\documentclass[aps,twocolumn,prd,showpacs,showkeys,preprintnumbers,superscriptaddress,nobibnotes,floatfix,longbibliography,notitlepage,nofootinbib]{revtex4-1}
\pdfoutput=1
\usepackage{amsmath}
\usepackage{amsfonts}
\usepackage{amssymb}
\usepackage{slashed}
\usepackage{mathrsfs}
\usepackage{graphicx}
\usepackage{xcolor}
\usepackage{xspace}
\usepackage{placeins}
\usepackage{booktabs}
\usepackage{soul}

\newcommand\colvec[3][]{\begin{pmatrix}\ifx\relax#1\relax\else#1\\\fi#2\\#3\end{pmatrix}}
\definecolor{darkmagenta}{rgb}{0.55, 0.0, 0.55}
\newcommand{\beq}{\begin{equation}}
\newcommand{\beqn}{\begin{eqnarray}}
\newcommand{\eeq}{\end{equation}}
\newcommand{\eeqn}{\end{eqnarray}}

\usepackage{mathtools,slashed}
\usepackage{array}
\usepackage{bigints}
\usepackage{tabularx}

\usepackage[colorlinks]{hyperref}
\hypersetup{
    colorlinks,
    linkcolor={red!50!black},
    citecolor={blue!50!black},
    urlcolor={blue!80!black}
}
\usepackage{multirow}
\usepackage{upgreek}
\usepackage[capitalise]{cleveref}
\usepackage{soul}
\usepackage{amsmath}

\def\mysection#1{{\bf #1.---} }

\begin{document}
\title{Direct Detection of the Millicharged Background}
\author{Ella Iles}
\affiliation{Department of Physics \& Trottier Space Institute, McGill University, Montr\'{e}al, Qu\'ebec H3A 2T8, Canada}

\author{Saniya Heeba}
\affiliation{Department of Physics \& Trottier Space Institute, McGill University, Montr\'{e}al, Qu\'ebec H3A 2T8, Canada}

\author{Katelin Schutz}
\affiliation{Department of Physics \& Trottier Space Institute, McGill University, Montr\'{e}al, Qu\'ebec H3A 2T8, Canada}

\begin{abstract}
\noindent 
We show that dark matter direct detection experiments are sensitive to the existence of particles with a small effective charge (for instance, via couplings to a kinetically mixed, low-mass dark photon). Our forecasts do \emph{not} depend on these particles comprising a significant fraction of the dark matter. Rather, these experiments are sensitive to the irreducible abundance produced in the early universe through the freeze-in mechanism. We find that ongoing and proposed direct detection experiments will have world-leading sensitivity to effective charges $Q\sim 10^{-12}$ across nine orders of magnitude in mass, corresponding to a dark matter sub-fraction as low as $\sim 10^{-3}$.
\end{abstract}

\maketitle
\mysection{Introduction}
The need for physics beyond the Standard Model (SM) -- including an explanation for the existence of dark matter (DM) -- strongly motivates the study of dark sectors that are only weakly coupled to the SM. One minimal example of such a sector is comprised of particles that are charged under a dark $U(1)^\prime$ gauge group (see e.g. Refs.~\cite{Boehm:2003hm, Pospelov:2007mp,Cheung:2007ut,Feldman:2007wj,Chun:2010ve}). Kinetic mixing between the dark and SM gauge bosons through $\kappa F^{\mu\nu}F^\prime_{\mu\nu}$ provides the weak coupling between this dark sector and the SM~\cite{Holdom:1985ag,Pospelov:2008zw,Fabbrichesi:2020wbt}. This dimension-four operator can originate from a range of UV scenarios, including string theory models where extended gauge sectors are ubiquitous~\cite{Dienes:1996zr,Abel:2008ai,Goodsell:2009xc,Shiu:2013wxa}. For ultralight $U(1)^\prime$ gauge bosons, the kinetic mixing can be rotated away such that the dark sector particles pick up a small effective electromagnetic charge (millicharge) under the SM $U(1)$ in the low-energy limit, 
\begin{equation}
\mathcal{L} \supset e Q\bar{\chi}\gamma_\mu\chi A^{\mu}+\bar{\chi}(i \slashed{\partial} -m_\chi)\chi,
\end{equation}
where the Dirac fermion $\chi$ is a millicharged particle (MCP) with dark charge $g_\chi$ and millicharge $Q = \kappa g_\chi/e$.

These theoretical considerations have made MCPs and dark photons a key benchmark for a range of collider and astronomical searches. Sub-MeV MCPs and dark photons can be produced copiously in stellar interiors and supernovae, impacting stellar evolution and other observables~\cite{An:2013yua,An:2013yfc,Redondo:2013lna,Giannotti:2015kwo,Chang:2018rso,Lasenby:2020goo,Dolan:2023cjs,Fung:2023euv} while colliders can employ missing energy and displaced vertex techniques to search for the production of MCPs and dark photons~\cite{Berlin:2018bsc,LDMX:2018cma,Ilten:2018crw,Alimena:2019zri}. Meanwhile, if MCPs are additionally assumed to comprise the observed DM of our Universe with $\Omega_\text{DM} h^2=0.12$~\cite{2020}, they also pose a key experimental target for ongoing DM direct detection searches including SENSEI~\cite{SENSEI:2023zdf} and DAMIC~\cite{DAMIC-M:2023gxo}, as well as the proposed OSCURA experiment~\cite{Oscura:2022vmi}. In fact, MCPs as DM are also one of the main theoretical motivations for new approaches to sub-MeV direct detection~\cite{Knapen:2017xzo}, for instance proposals involving direct deflection~\cite{Berlin_2020}, polar materials (GaAs and Al$_2$O$_3$)~\cite{Knapen:2017ekk,polarmaterials:2018bjn} (which can include a multiphonon response~\cite{Campbell-Deem:2022fqm,priv} detectable with SPICE~\cite{derenzosnowmass2021}), Dirac materials (ZrTe$_5$, Yb$_3$PbO, and BNQ-TTF)~\cite{Hochberg:2017wce,Geilhufe_2020DiracMaterials}, doped semiconductors~\cite{Du:2022dxf}, or superconductors (Al SC)~\cite{Hochberg:2021pkt,Knapen:2021run}. 

\begin{figure}[t!]
        \centering
        \includegraphics[width = 0.49\textwidth]{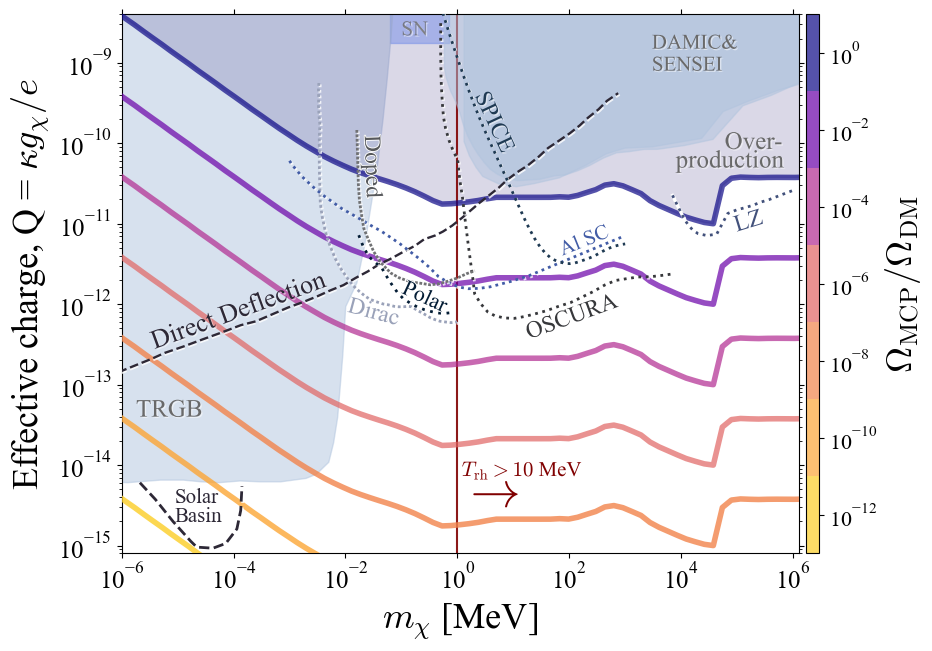}
        \vspace{-0.8cm}
        \caption{Experimental reach of searches for MCPs. The solid lines represent the freeze-in relic abundance of MCPs normalized to the DM abundance. The shaded regions represent constraints from the TRGB~\cite{Fung:2023euv}, SENSEI~\cite{SENSEI:2023zdf},  DAMIC~\cite{DAMIC-M:2023gxo}, and supernova 1987A~\cite{Chang:2018rso}. The dotted lines correspond to proposed direct detection searches for MCP scattering, while the dashed lines correspond to the forecast sensitivity of direct deflection~\cite{Berlin_2020} (including a solar MCP basin~\cite{Berlin:2021kcm}). Each of the direct detection lines were shifted taking into account the variable MCP abundance in the $Q\,$--$\,m_\chi$ plane. 
        }
        \vspace{-0.4cm}
        \label{summary}
\end{figure}

As summarized in Fig.~\ref{summary}, in this \textit{Letter} we point out that if MCPs exist as particle states in the spectrum, then some MCP abundance will inevitably be produced in the early universe, leading to an irreducible cosmic MCP background. In particular, MCPs can be produced non-thermally via freeze-in, where annihilations and decays of SM states generate a stable density of MCPs. MCP freeze-in is IR-dominated and proceeds until the process is either kinematically disallowed or until the abundance of SM initial states is depleted by the expansion of the universe. For $Q\sim10^{-11}$, this slow accumulation of MCPs can account for the observed DM abundance \cite{JHEP,Chu:2011be,FIMPdawn,Dvorkin_2019}, providing a target for DM direct detection experiments. At even larger values of $Q$, the irreducible MCP abundance is larger than the DM abundance, effectively ruling out the parameter space~\cite{Gan:2023jbs}. Meanwhile, for lower values of the millicharge, the MCP background constitutes a small sub-component of the DM. We note that similar arguments have previously been applied to axions and sterile neutrinos~\cite{Langhoff_2022}, as well as real scalars~\cite{DEramo:2024lsk}, resulting in stringent constraints based on their freeze-in abundance. Notably, axion freeze-in can be UV-dominated and is sensitive to the reheating temperature of the Universe, whereas sub-MeV MCP freeze-in happens roughly at the same time as Big Bang Nucelosynthesis, regardless of the reheating temperature. 

Strikingly, as summarized in Fig.~\ref{summary}, a MCP subcomponent of DM will still be accessible with any DM direct detection searches that have sensitivity to couplings below the freeze-in target, resulting in general bounds on the existence of MCPs regardless of whether they are the DM of our Universe. We find that traditional direct detection searches for MCP-SM scattering (whose sensitivity scales linearly with the MCP density) will be able to constrain a fractional MCP abundance as low as $\sim 10^{-3}$ for keV-TeV masses. Meanwhile, sub-keV MCPs will be most readily accessible to a direct deflection approach (whose sensitivity scales like the MCP density squared), potentially reaching fractional MCP densities eight orders of magnitude below the DM density.
\label{sec:intro}

\mysection{Irreducible MCP Density} For sub-MeV masses we extend the freeze-in calculation of Ref. \cite{Dvorkin_2019} to arbitrary abundances, including sub-keV MCP masses. The production of sub-MeV MCPs occurs via two main interactions: the annihilation of electron-positron pairs, $e^+e^- \to \bar{\chi}\chi$ \cite{DDe+e-,DDxenon} and the in-medium decay of plasmons~\cite{Dvorkin_2019,Berlin:2022hmt}. The abundance of MCPs can then be calculated by solving the Boltzmann equation,
\begin{equation}
    sHx\frac{dY_\chi}{dx}=2\left(n_e^2 \langle\sigma v\rangle_{e^+e^-\rightarrow\chi\bar{\chi}}+ n_{\gamma^*}\langle\Gamma\rangle_{\gamma^*\rightarrow\chi\bar{\chi}}\right)
    \label{Boltzmann1}
\end{equation}
where $Y_\chi=n_\chi/s$ is the comoving MCP number density, $H$ is the Hubble rate, $s$ is the entropy density, $x = m_\chi/T$, and the two terms on the right are the thermally-averaged annihilation and plasmon decay rates, respectively. The factor of 2 accounts for the abundance of both $\chi$ and $\bar{\chi}$. We do not include a term that captures the backreaction of MCP annihilation or coalescence into SM particles, as we have explicitly checked that the number density of MCPs is always highly suppressed compared to that of SM particles during freeze-in. 

MCP production via annihilation occurs until $T \sim {\rm max} (m_e,\,m_\chi)$, corresponding to temperatures when the electrons either freeze-out of the thermal bath or lack sufficient energy to produce MCPs. On the other hand, plasmon decay remains active until the effective plasmon mass (which is similar to the plasma frequency $\omega_p$) drops below twice the MCP mass. At early times $T\gtrsim1$~MeV, the plasma is relativistic so $\omega_p \approx e T/3$, whereas for $T\lesssim1$~MeV the plasma frequency depends primarily on the ambient electron density $\omega_p^2 \approx e^2 n_e/m_e$. For sufficiently low-mass MCPs, this means that production occurs in a regime where the chemical potential of electrons $\mu_e$ cannot be neglected. We therefore include the electron chemical potential in our calculations, which we obtain by equating the net electron density to the baryon density at all times. At late times when the plasma is nonrelativistic, we find that the electron chemical potential can be approximated as 
\begin{equation}
\mu_e=T\,\sinh^{-1}\left[\eta\frac{\zeta(3)}{\pi^2}T^3e^{\frac{m_e}{T}}\left(\frac{m_eT}{2\pi}\right)^{-3/2}\right],
\end{equation}
in agreement with other treatments in the literature~\cite{Thomas_2020}. 
 
\begin{figure}[t!]
    \centering
    \includegraphics[width=0.49\textwidth]{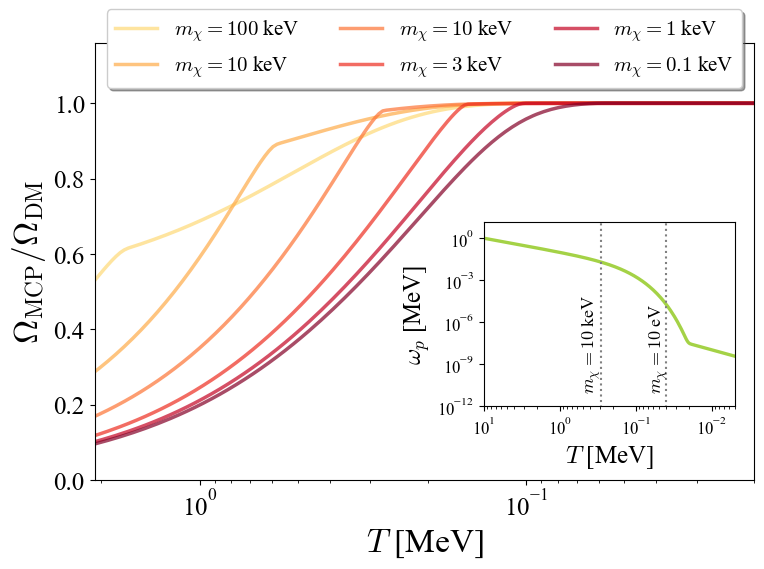}
    \vspace{-0.5cm}
    \caption{ Evolution of the relic MCP abundance for several MCP masses with values of $Q$ chosen such that $\Omega_\text{MCP}=\Omega_\text{DM}$. 
    The curves converge at lower MCP masses, where plasmon decay becomes inefficient at $T\sim0.1$~MeV primarily due to the exponentially falling plasma frequency (shown in the inset) rather than due to the kinematic threshold $\omega_p > 2 m_\chi$ (illustrated with vertical dotted lines for two choices of $m_\chi$). MCP production is thus complete by $T\sim 0.1$~MeV at the very latest, regardless of the MCP mass.
    }
    \label{reaction rates}
    \vspace{-0.4cm}
\end{figure}

Although we solve Eq.~\eqref{Boltzmann1} numerically, including the effects of finite temperature and chemical potential, the relative contribution of the two channels as a function of temperature can be understood using simple scaling arguments. The rates for both annihilation and plasmon decay scale as $\sim T$ whereas the Hubble rate scales as $\sim T^2$, indicating that freeze-in is most active at the lowest temperatures that are kinematically accessible (or that have a sufficient abundance of SM initial states). It is therefore possible to approximate freeze-in as happening at a particular temperature to compute the yield using the relevant collision term in the Boltzmann equation. For $e^+e^-$ annihilation, $\langle \sigma v \rangle n_e^2 \sim e^4 Q^2 T^4$ and the lowest temperature where the process is active is $T\sim \max(m_\chi, m_e)$, giving a comoving MCP abundance that scales as 
\begin{equation}
    Y_\chi^{e^+e^-}= 1\times10^{-9}\left(\frac{Q}{10^{-12}}\right)^2\left(\frac{1\text{ MeV}}{{\rm max} (m_\chi,\,m_e)}\right).
    \label{Ye+e-1}
\end{equation}
On the other hand, the plasmon decay term scales as $\langle \Gamma\rangle_{\gamma^*\rightarrow\chi\bar{\chi}} n_{\gamma^*}\sim e^2 Q^2 \omega_p T^3$. As the temperature of the universe decreases, the plasma frequency decreases with a non-trivial scaling (shown as an inset plot in Figure \ref{reaction rates}), quenching the plasmon decay channel at a characteristic temperature that depends on $m_\chi$. For MCPs above a mass of $m_\chi \gtrsim 10$~keV, plasmon decay ceases at the kinematic threshold, $\omega_p(T) \sim 2 m_\chi$. For a $\sim 10$~keV MCP, this threshold occurs at $T\sim$~0.2~MeV, when the plasma is still fairly relativistic. This results in a comoving abundance,
\begin{equation}
Y_\chi^{\gamma^*}=2\times10^{-10}\left(\frac{Q}{10^{-12}}\right)^2  \left(\frac{1\text{ MeV}}{m_\chi}\right) \quad \text{if } m_\chi \gtrsim 10 \text{ keV}.\label{Ygamma1}
\end{equation} 
For MCPs with lower masses, MCP production is still kinematically allowed during
electron freeze-out, which results in an exponential suppression of the plasma frequency. In this case, the electron depletion is primarily responsible for quenching plasmon decay. The production of the MCP background is therefore completed by $T\sim 0.1$ MeV at the very latest, just before the strong exponential suppression of the electron density, regardless of MCP mass or production channel. This results in a comoving MCP abundance proportional to
\begin{equation}
Y_\chi^{\gamma^*}=3\times10^{-8}\left(\frac{Q}{10^{-12}}\right)^2 \quad \text{if } m_\chi \lesssim 10\text{ keV},  \label{Ygamma1b}
\end{equation}
in agreement with Ref.~\cite{Berlin:2022hmt}. Eqs.~\eqref{Ye+e-1}-\eqref{Ygamma1b} make it apparent that when $m_\chi \gtrsim 100$~keV, the electron-positron annihilation rate dominates over plasmon decay. Consequently, the total MCP abundance, $\Omega_{\rm MCP} = m_\chi(Y_{e^+e^-} + Y_{\gamma^*})s_0/\rho_0^c$ scales as $\Omega_{\rm MCP}\sim Q^2$ for $m_\chi \gtrsim 10$~keV and as $\Omega_{\rm MCP}\sim Q^2 m_\chi$ for $m_\chi \lesssim 10$~keV, reproducing the scaling seen in Fig.~\ref{summary}. 

For MCP masses $m_\chi \gg 1$ MeV, freeze-in production primarily happens through SM fermions $f\bar{f}\to \chi\bar{\chi}$, as calculated in Ref.~\cite{Chu:2011be}. We ignore the effect of Fermi-Dirac statistics, which can shift lines of constant relic abundance by $O(10\%)$ \cite{Heeba:2023bik}. Since the abundance through these channels is proportional to $Q^2$, one can straightforwardly generalize these results to arbitrary values of $Q$ with a simple rescaling, $\Omega_{\rm MCP} = \Omega_{\rm DM} (Q/Q_{\rm DM})^2$, where $Q_{\rm DM}$ is the millicharge required to produce the observed DM abundance $\Omega_{\rm DM}$. Note that at energies near or above the electroweak scale, MCPs additionally couple to the SM $Z$-boson~\cite{Evans:2017kti}. Consequently, MCPs in the mass range $1 {\rm GeV} \lesssim m_\chi \lesssim m_Z/2$ are produced through the decay channel $Z\rightarrow \chi\bar{\chi}$, resulting in the drop in $Q$ required for a fixed abundance as computed in Ref.~\cite{Heeba:2023bik}. Note that the same finite-temperature effects that result in modified photon mixing properties can also affect the $Z$. However, these corrections are expected to be very small~\cite{PhysRevD.55.6253,Heeba:2019jho}, and therefore should not substantially impact our result. For $m_\chi \gtrsim 200\, {\rm GeV}$, production will dominantly happen before the electroweak phase transition, where finite-temperature effects may require a more careful treatment beyond the scope of this work. 

Our calculations of the MCP abundance are subject to the caveat that the MCPs cannot be substantially depleted after their production. Notably, the presence of an ultralight $A^\prime$ in the spectrum implies that the MCP abundance can be depleted through $\chi\bar{\chi}\to A^\prime A^\prime$. Since this process scales as $g_\chi^4$, it can potentially be efficient despite the sub-thermal abundance of MCPs. Ensuring that this process is always inefficient implies that our calculations are valid when $g_\chi < 3\times 10^{-4}\times  (m_\chi / 1\,\text{MeV})^{1/2} (\Omega_\text{DM}/\Omega_\text{MCP})^{1/4}$. Small values of the dark gauge coupling, combined with the freeze-in values of $Q$ and implied kinetic mixing parameter $\kappa$ may be bolstered by UV considerations~\cite{Reece:2018zvv,Gherghetta:2019coi}. This bound on the dark gauge coupling is also similar to the one obtained through constraints on self-interacting DM (SIDM) that apply when MCPs are assumed to make up all of the dark matter \cite{Tulin:2017ara,Kummer:2019yrb}. We finally note that it is possible to entirely avoid the depletion of MCPs by considering the the ``pure'' millicharge case where the MCPs have a small SM hypercharge and thus have a direct coupling to the SM photon. In this case, the only annihilation channel $\chi \bar{\chi}\rightarrow \gamma \gamma $ is suppressed like $Q^4$ and is therefore safely neglected in the entire parameter space. 

\label{sec:MCP}

\mysection{Experimental Reach} 
By accounting for the variable relic abundance of MCPs in the $Q-m_\chi$ plane, we can recast existing forecasts and constraints that assume MCPs are all of the DM into constraints and forecasts on the mere existence of MCPs. Smaller abundances of MCPs correspond to reduced overall experimental sensitivity due to the reduced flux of particles passing through the detector. We assume that MCPs produced in the early universe have the same gravitational clustering dynamics as the DM, which would imply that the local MCP density corresponds to the rescaled DM density, 
\begin{equation}
    \rho_\text{MCP}(Q,m_\chi)=\rho_\text{DM}\frac{\Omega_\text{MCP}(Q,m_\chi)}{\Omega_\text{DM}}
    \label{eq:MCPDMratio}
\end{equation}
where $\rho_\text{DM}\sim 0.3$~ GeV/cm$^3$ is the local mass density of DM in the Milky Way \cite{Read:2014qva,deSalas:2020hbh}. We also assume the MCP background has the same velocity distribution as the DM for similar reasons (most often, direct detection analyses are performed assuming a Maxwellian distribution). To be consistent with these assumptions about the local MCP density and phase space, we only consider MCP masses above $\sim 1$~eV, since even lighter MCPs would have still have substantial free-streaming velocities at the present day (MCPs lighter than $\sim10^{-4}$~eV would still be relativistic).

Ongoing and proposed sub-GeV direct detection experiments provide a key avenue for constraining the irreducible MCP background at intermediate MCP masses. Existing analyses and forecasts for these experiments place a bound on the electron scattering cross-section, 
\begin{align}
    \overline{\sigma}_e = \frac{16\pi \alpha^2 Q^2  \mu_{\chi e}^2}{(\alpha m_e)^4}\,,
\end{align}
as a function of $m_\chi$ (where $\mu_{\chi e}$ is the MCP-electron reduced mass) under the assumption that MCPs make all of the DM. The event rate at these experiments, which determines their sensitivity, scales as $R \propto \rho_{\rm{DM}} \overline{\sigma}_e/m_{\rm DM}$. For the MCP background, we can therefore interpret existing limits on $\overline{\sigma}_e$ as limits on $  \overline{\sigma}_e \times (\rho_{\rm{MCP}}/\rho_{\rm DM})$, which can be straightforwardly projected onto the $Q-m_\chi$ plane. 

For heavier MCP masses, the experimental program targeting weakly interacting massive particles (WIMPs) via nuclear recoils can be repurposed to search for MCPs. In the mass range $m_\chi \gtrsim {\rm GeV}$, MCPs can be detected via their interaction with the protons in the target nucleus. Generally, bounds on the DM-proton cross section assume a heavy $A^\prime$ such that $\sigma_p$ is suppressed by $m_{A^\prime}^{-4}$. However these bounds can be reinterpreted for the case of an ultralight $A^\prime$ or a ``pure" millicharge, resulting in an enhancement 
at low recoil energies,
\begin{equation}
    \sigma_p=\frac{16\pi \alpha^2Q^2 \mu_{\chi p}^2}{(2m_NE_R)^2}
    \label{eq:sigman}
\end{equation}
where $\mu_{\chi p}^2$ is the MCP-proton reduced mass, $m_N$ is the mass of the target nucleus, and $E_R$ is the recoil energy. Despite the fact that the differential event rate for a fixed DM mass is different for heavy and light mediators, Ref.~\cite{Hambye:2018dpi} pointed out that the two differential event rates can be similar for \textit{distinct} DM masses. Therefore constraints assuming a heavy mediator can be reinterpreted as constraints on MCPs (including the fractional abundance of  Eq.~\eqref{eq:MCPDMratio}) by ensuring that the differential event rates are the same in the two cases~\cite{Hambye:2018dpi}. 

New detection strategies, such as direct deflection~\cite{Berlin_2020}, are necessary to access sub-keV MCP parameter space. Notably, it would be challenging to detect energy deposition from single-particle scattering at the $\sim 1$~meV scale, corresponding to a keV-mass particle orbiting in the MW. Direct deflection assumes that the ``MCP wind" passing through earth acts like a continuum, and the signal arises due to the collective deflection of the entire wind in an oscillating background electric field, which sources an alternating millicharge electric field downstream of the deflector. The millicharge-induced electric field scales as $E_\chi \sim  \rho_\text{MCP}Q^2/m_\chi^2$, and the signal power integrated over the shielded detector region scales as $E_\chi^2 \sim \rho_\text{MCP}^2 (Q/m_\chi)^4$. We can thus appropriately rescale the sensitivity of DM searches using direct deflection to searches for MCP states in the spectrum. 
The same direct deflection setup would also be sensitive to the MCPs generated by the Sun that are trapped in the Solar basin~\cite{Berlin:2021kcm}. The sensitivity to the Solar basin MCPs is stronger due to the higher local number density and slower kinematics of basin-generated MCPs compared to frozen-in MCPs.

\label{sec:constraints}

\mysection{Astrophysical and Cosmological Constraints} The {overproduction} bound shown in Fig. \ref{summary} can be understood as the parameter space where the abundance of MCPs produced via freeze-in would be larger than the observed abundance of DM, as discussed in Ref.~\cite{Gan:2023jbs} (or e.g. Ref.~\cite{Lebedev:2022cic} in the case of scalars). {We note that this overproduction bound is alleviated for $g_\chi \gtrsim 10^{-4}$ when MCP annihilation to dark photons becomes efficient, which opens up the parameter space.} Furthermore, throughout this work, we have assumed that the reheating temperature of the Universe is much larger than any of the other relevant temperature scales. However, non-standard early-Universe scenarios could result in lower reheating temperatures that are constrained by observation to be at least $\sim10$~MeV~\cite{Hasegawa:2019jsa,deSalas:2015glj,Hannestad:2004px}. The vertical line in Fig.~\ref{summary} shows the approximate parameter space where we expect this assumption to affect our calculation. Lowering the reheat temperature could alleviate the overproduction bound for heavy MCPs, as discussed in Ref.~\cite{Gan:2023jbs}. Additionally, frozen-in MCPs never have a thermal abundance and thus never possess substantial entropy, indicating that they do not impact the measured value of $N_\text{eff}$, in contrast to freeze-out scenarios~\cite{Gan:2023jbs}. 

There are many methods to constrain MCPs, for instance from anomalous energy loss from stars and supernovae~\cite{Fung:2023euv,Chang:2018rso}. Direct detection searches would provide complementary constraints in the low-mass parameter space, which is particularly important since stellar magnetic fields could trap MCPs in stars, in violation of the assumptions typically made in setting stellar energy-loss constraints. Dedicated work is required to understand magnetic trapping in stars, but studies of the Sun suggest that parameter space above $Q\gtrsim 10^{-11}$ could be impacted~\cite{Vinyoles:2015khy}. 
Additionally, if MCPs are assumed to make up all of the DM, they can leave imprints in many astrophysical environments. For instance, the initial phase space distribution from freeze-in production can result in suppressed structure formation on small scales~\cite{Dvorkin:2020xga}. Nontrivial MCP magneto-hydrodynamics may also affect the internal properties of galaxies and galaxy clusters \cite{Cruz:2022otv,Stebbins:2019xjr,Lasenby:2020rlf}, particularly accounting for ambient magnetic fields and supernova remnant shocks that may alter the local density and phase-space distribution of MCPs in our Galaxy~\cite{Chuzhoy:2008zy,Foot:2010yz,Dunsky:2018mqs,Li:2020wyl,Berlin:2023zpn}. However, fully characterizing these dark plasma effects (including possible screening due to a finite dark photon mass) remains a challenging problem. The scattering of frozen-in MCPs with baryons can also affect the cosmic microwave background~\cite{Dvorkin:2020xga} and
the 21~cm global signal~\cite{PhysRevLett.121.011102,Barkana:2022hko,Barkana:2018qrx,Liu:2019knx}. We expect that the these effects will be significantly diminished if MCPs are a small fraction of the total DM abundance. For instance, it has been shown that MCP fractions less than 0.4\% cannot be constrained from DM-baryon scattering in the CMB ~\cite{Boddy:2018wzy} (assuming cold initial conditions). We finally note that fermionic MCPs will not be subject to bounds from Pauli blocking in dwarf galaxies (see e.g. Ref.~\cite{Alvey:2020xsk}) if they only constitute a small 
DM fraction. 

\mysection{Conclusions}
MCPs arise as parts of minimal  dark sectors that are weakly coupled to the SM. If MCPs exist in nature, they may not constitute a substantial fraction of the DM. Regardless, in this work we have shown that DM direct detection experiments are sensitive to the existence of MCPs (even if they are a tiny sub-fraction of DM) due to the irreducible MCP density produced in the early universe. In the context of freeze-in production, smaller couplings result in smaller DM fractions (this is in contrast to the freeze-out mechanism, where the opposite is true). Thus, any experiment that has sensitivity below the traditional freeze-in target can probe the existence of a frozen-in MCP background. Notably, we find that direct detection experiments will be able to achieve world-leading sensitivity to MCPs with masses spanning nine orders of magnitude. This sensitivity to dark sectors further motivates the future direct detection program. 

Our results are conservative in the sense that we do not include channels for MCP production outside of freeze-in, which would only increase the MCP density and strengthen the experimental reach. For instance, fully non-thermal production channels like misalignment may also contribute substantially to the ambient density of low-mass MCPs~\cite{Bogorad:2021uew}. Our results can also be extended to consider other thermal histories, for instance with an initial dark sector abundance~\cite{Fernandez:2021iti}.
\label{sec:conclusions}

\mysection{Acknowledgements}
It is a pleasure to thank Asher Berlin, David Dunsky, Rouven Essig, Felix Kahlhoefer, Yoni Kahn, Tongyan Lin, Nadav Outmezguine, and Nick Rodd for useful conversations and comments on the manuscript. EI was supported in part by a Natural Sciences and Engineering Research Council of Canada (NSERC) Undergraduate Student Research Award USRA-594131-2024. SH was supported in part by the Canadian Institute of Particle Physics Connect Fellowship. EI, SH, and KS acknowledge support from a NSERC Subatomic Physics Discovery Grant, and from the Canada Research Chairs program. KS thanks the Kavli Institute for Theoretical Physics (supported by grant NSF PHY-2309135) for their hospitality during the completion of this work. This analysis made use of \texttt{Numpy} \cite{harris2020array}, \texttt{Scipy} \cite{virtanen2020scipy}, \texttt{Matplotlib} \cite{hunter2007matplotlib}, \texttt{WebPlotDigitizer} \cite{rohatgi2018webplotdigitizer}, \texttt{Jupyter} \cite{soton403913}  and \texttt{Mathematica}~\cite{wolfram2003mathematica}.
\bibliography{citations}

\begin{thebibliography}{91}%
\makeatletter
\providecommand \@ifxundefined [1]{%
 \@ifx{#1\undefined}
}%
\providecommand \@ifnum [1]{%
 \ifnum #1\expandafter \@firstoftwo
 \else \expandafter \@secondoftwo
 \fi
}%
\providecommand \@ifx [1]{%
 \ifx #1\expandafter \@firstoftwo
 \else \expandafter \@secondoftwo
 \fi
}%
\providecommand \natexlab [1]{#1}%
\providecommand \enquote  [1]{``#1''}%
\providecommand \bibnamefont  [1]{#1}%
\providecommand \bibfnamefont [1]{#1}%
\providecommand \citenamefont [1]{#1}%
\providecommand \href@noop [0]{\@secondoftwo}%
\providecommand \href [0]{\begingroup \@sanitize@url \@href}%
\providecommand \@href[1]{\@@startlink{#1}\@@href}%
\providecommand \@@href[1]{\endgroup#1\@@endlink}%
\providecommand \@sanitize@url [0]{\catcode `\\12\catcode `\$12\catcode
  `\&12\catcode `\#12\catcode `\^12\catcode `\_12\catcode `\%12\relax}%
\providecommand \@@startlink[1]{}%
\providecommand \@@endlink[0]{}%
\providecommand \url  [0]{\begingroup\@sanitize@url \@url }%
\providecommand \@url [1]{\endgroup\@href {#1}{\urlprefix }}%
\providecommand \urlprefix  [0]{URL }%
\providecommand \Eprint [0]{\href }%
\providecommand \doibase [0]{http://dx.doi.org/}%
\providecommand \selectlanguage [0]{\@gobble}%
\providecommand \bibinfo  [0]{\@secondoftwo}%
\providecommand \bibfield  [0]{\@secondoftwo}%
\providecommand \translation [1]{[#1]}%
\providecommand \BibitemOpen [0]{}%
\providecommand \bibitemStop [0]{}%
\providecommand \bibitemNoStop [0]{.\EOS\space}%
\providecommand \EOS [0]{\spacefactor3000\relax}%
\providecommand \BibitemShut  [1]{\csname bibitem#1\endcsname}%
\let\auto@bib@innerbib\@empty
\bibitem [{\citenamefont {Boehm}\ and\ \citenamefont
  {Fayet}(2004)}]{Boehm:2003hm}%
  \BibitemOpen
  \bibfield  {author} {\bibinfo {author} {\bibfnamefont {C.}~\bibnamefont
  {Boehm}}\ and\ \bibinfo {author} {\bibfnamefont {Pierre}\ \bibnamefont
  {Fayet}},\ }\bibfield  {title} {\enquote {\bibinfo {title} {{Scalar dark
  matter candidates}},}\ }\href {\doibase 10.1016/j.nuclphysb.2004.01.015}
  {\bibfield  {journal} {\bibinfo  {journal} {Nucl. Phys. B}\ }\textbf
  {\bibinfo {volume} {683}},\ \bibinfo {pages} {219--263} (\bibinfo {year}
  {2004})},\ \Eprint {http://arxiv.org/abs/hep-ph/0305261}
  {arXiv:hep-ph/0305261} \BibitemShut {NoStop}%
\bibitem [{\citenamefont {Pospelov}\ \emph {et~al.}(2008)\citenamefont
  {Pospelov}, \citenamefont {Ritz},\ and\ \citenamefont
  {Voloshin}}]{Pospelov:2007mp}%
  \BibitemOpen
  \bibfield  {author} {\bibinfo {author} {\bibfnamefont {Maxim}\ \bibnamefont
  {Pospelov}}, \bibinfo {author} {\bibfnamefont {Adam}\ \bibnamefont {Ritz}}, \
  and\ \bibinfo {author} {\bibfnamefont {Mikhail~B.}\ \bibnamefont
  {Voloshin}},\ }\bibfield  {title} {\enquote {\bibinfo {title} {{Secluded WIMP
  Dark Matter}},}\ }\href {\doibase 10.1016/j.physletb.2008.02.052} {\bibfield
  {journal} {\bibinfo  {journal} {Phys. Lett. B}\ }\textbf {\bibinfo {volume}
  {662}},\ \bibinfo {pages} {53--61} (\bibinfo {year} {2008})},\ \Eprint
  {http://arxiv.org/abs/0711.4866} {arXiv:0711.4866 [hep-ph]} \BibitemShut
  {NoStop}%
\bibitem [{\citenamefont {Cheung}\ and\ \citenamefont
  {Yuan}(2007)}]{Cheung:2007ut}%
  \BibitemOpen
  \bibfield  {author} {\bibinfo {author} {\bibfnamefont {Kingman}\ \bibnamefont
  {Cheung}}\ and\ \bibinfo {author} {\bibfnamefont {Tzu-Chiang}\ \bibnamefont
  {Yuan}},\ }\bibfield  {title} {\enquote {\bibinfo {title} {{Hidden fermion as
  milli-charged dark matter in Stueckelberg Z- prime model}},}\ }\href
  {\doibase 10.1088/1126-6708/2007/03/120} {\bibfield  {journal} {\bibinfo
  {journal} {JHEP}\ }\textbf {\bibinfo {volume} {03}},\ \bibinfo {pages} {120}
  (\bibinfo {year} {2007})},\ \Eprint {http://arxiv.org/abs/hep-ph/0701107}
  {arXiv:hep-ph/0701107} \BibitemShut {NoStop}%
\bibitem [{\citenamefont {Feldman}\ \emph {et~al.}(2007)\citenamefont
  {Feldman}, \citenamefont {Liu},\ and\ \citenamefont {Nath}}]{Feldman:2007wj}%
  \BibitemOpen
  \bibfield  {author} {\bibinfo {author} {\bibfnamefont {Daniel}\ \bibnamefont
  {Feldman}}, \bibinfo {author} {\bibfnamefont {Zuowei}\ \bibnamefont {Liu}}, \
  and\ \bibinfo {author} {\bibfnamefont {Pran}\ \bibnamefont {Nath}},\
  }\bibfield  {title} {\enquote {\bibinfo {title} {{The Stueckelberg Z-prime
  Extension with Kinetic Mixing and Milli-Charged Dark Matter From the Hidden
  Sector}},}\ }\href {\doibase 10.1103/PhysRevD.75.115001} {\bibfield
  {journal} {\bibinfo  {journal} {Phys. Rev. D}\ }\textbf {\bibinfo {volume}
  {75}},\ \bibinfo {pages} {115001} (\bibinfo {year} {2007})},\ \Eprint
  {http://arxiv.org/abs/hep-ph/0702123} {arXiv:hep-ph/0702123} \BibitemShut
  {NoStop}%
\bibitem [{\citenamefont {Chun}\ \emph {et~al.}(2011)\citenamefont {Chun},
  \citenamefont {Park},\ and\ \citenamefont {Scopel}}]{Chun:2010ve}%
  \BibitemOpen
  \bibfield  {author} {\bibinfo {author} {\bibfnamefont {Eung~Jin}\
  \bibnamefont {Chun}}, \bibinfo {author} {\bibfnamefont {Jong-Chul}\
  \bibnamefont {Park}}, \ and\ \bibinfo {author} {\bibfnamefont {Stefano}\
  \bibnamefont {Scopel}},\ }\bibfield  {title} {\enquote {\bibinfo {title}
  {{Dark matter and a new gauge boson through kinetic mixing}},}\ }\href
  {\doibase 10.1007/JHEP02(2011)100} {\bibfield  {journal} {\bibinfo  {journal}
  {JHEP}\ }\textbf {\bibinfo {volume} {02}},\ \bibinfo {pages} {100} (\bibinfo
  {year} {2011})},\ \Eprint {http://arxiv.org/abs/1011.3300} {arXiv:1011.3300
  [hep-ph]} \BibitemShut {NoStop}%
\bibitem [{\citenamefont {Holdom}(1986)}]{Holdom:1985ag}%
  \BibitemOpen
  \bibfield  {author} {\bibinfo {author} {\bibfnamefont {Bob}\ \bibnamefont
  {Holdom}},\ }\bibfield  {title} {\enquote {\bibinfo {title} {{Two U(1)'s and
  Epsilon Charge Shifts}},}\ }\href {\doibase 10.1016/0370-2693(86)91377-8}
  {\bibfield  {journal} {\bibinfo  {journal} {Phys. Lett. B}\ }\textbf
  {\bibinfo {volume} {166}},\ \bibinfo {pages} {196--198} (\bibinfo {year}
  {1986})}\BibitemShut {NoStop}%
\bibitem [{\citenamefont {Pospelov}(2009)}]{Pospelov:2008zw}%
  \BibitemOpen
  \bibfield  {author} {\bibinfo {author} {\bibfnamefont {Maxim}\ \bibnamefont
  {Pospelov}},\ }\bibfield  {title} {\enquote {\bibinfo {title} {{Secluded U(1)
  below the weak scale}},}\ }\href {\doibase 10.1103/PhysRevD.80.095002}
  {\bibfield  {journal} {\bibinfo  {journal} {Phys. Rev. D}\ }\textbf {\bibinfo
  {volume} {80}},\ \bibinfo {pages} {095002} (\bibinfo {year} {2009})},\
  \Eprint {http://arxiv.org/abs/0811.1030} {arXiv:0811.1030 [hep-ph]}
  \BibitemShut {NoStop}%
\bibitem [{\citenamefont {Fabbrichesi}\ \emph {et~al.}(2020)\citenamefont
  {Fabbrichesi}, \citenamefont {Gabrielli},\ and\ \citenamefont
  {Lanfranchi}}]{Fabbrichesi:2020wbt}%
  \BibitemOpen
  \bibfield  {author} {\bibinfo {author} {\bibfnamefont {Marco}\ \bibnamefont
  {Fabbrichesi}}, \bibinfo {author} {\bibfnamefont {Emidio}\ \bibnamefont
  {Gabrielli}}, \ and\ \bibinfo {author} {\bibfnamefont {Gaia}\ \bibnamefont
  {Lanfranchi}},\ }\bibfield  {title} {\enquote {\bibinfo {title} {{The Dark
  Photon}},}\ }\href {\doibase 10.1007/978-3-030-62519-1} {\  (\bibinfo {year}
  {2020}),\ 10.1007/978-3-030-62519-1},\ \Eprint
  {http://arxiv.org/abs/2005.01515} {arXiv:2005.01515 [hep-ph]} \BibitemShut
  {NoStop}%
\bibitem [{\citenamefont {Dienes}\ \emph {et~al.}(1997)\citenamefont {Dienes},
  \citenamefont {Kolda},\ and\ \citenamefont {March-Russell}}]{Dienes:1996zr}%
  \BibitemOpen
  \bibfield  {author} {\bibinfo {author} {\bibfnamefont {Keith~R.}\
  \bibnamefont {Dienes}}, \bibinfo {author} {\bibfnamefont {Christopher~F.}\
  \bibnamefont {Kolda}}, \ and\ \bibinfo {author} {\bibfnamefont {John}\
  \bibnamefont {March-Russell}},\ }\bibfield  {title} {\enquote {\bibinfo
  {title} {{Kinetic mixing and the supersymmetric gauge hierarchy}},}\ }\href
  {\doibase 10.1016/S0550-3213(97)00173-9} {\bibfield  {journal} {\bibinfo
  {journal} {Nucl. Phys. B}\ }\textbf {\bibinfo {volume} {492}},\ \bibinfo
  {pages} {104--118} (\bibinfo {year} {1997})},\ \Eprint
  {http://arxiv.org/abs/hep-ph/9610479} {arXiv:hep-ph/9610479} \BibitemShut
  {NoStop}%
\bibitem [{\citenamefont {Abel}\ \emph {et~al.}(2008)\citenamefont {Abel},
  \citenamefont {Goodsell}, \citenamefont {Jaeckel}, \citenamefont {Khoze},\
  and\ \citenamefont {Ringwald}}]{Abel:2008ai}%
  \BibitemOpen
  \bibfield  {author} {\bibinfo {author} {\bibfnamefont {S.~A.}\ \bibnamefont
  {Abel}}, \bibinfo {author} {\bibfnamefont {M.~D.}\ \bibnamefont {Goodsell}},
  \bibinfo {author} {\bibfnamefont {J.}~\bibnamefont {Jaeckel}}, \bibinfo
  {author} {\bibfnamefont {V.~V.}\ \bibnamefont {Khoze}}, \ and\ \bibinfo
  {author} {\bibfnamefont {A.}~\bibnamefont {Ringwald}},\ }\bibfield  {title}
  {\enquote {\bibinfo {title} {{Kinetic Mixing of the Photon with Hidden U(1)s
  in String Phenomenology}},}\ }\href {\doibase 10.1088/1126-6708/2008/07/124}
  {\bibfield  {journal} {\bibinfo  {journal} {JHEP}\ }\textbf {\bibinfo
  {volume} {07}},\ \bibinfo {pages} {124} (\bibinfo {year} {2008})},\ \Eprint
  {http://arxiv.org/abs/0803.1449} {arXiv:0803.1449 [hep-ph]} \BibitemShut
  {NoStop}%
\bibitem [{\citenamefont {Goodsell}\ \emph {et~al.}(2009)\citenamefont
  {Goodsell}, \citenamefont {Jaeckel}, \citenamefont {Redondo},\ and\
  \citenamefont {Ringwald}}]{Goodsell:2009xc}%
  \BibitemOpen
  \bibfield  {author} {\bibinfo {author} {\bibfnamefont {Mark}\ \bibnamefont
  {Goodsell}}, \bibinfo {author} {\bibfnamefont {Joerg}\ \bibnamefont
  {Jaeckel}}, \bibinfo {author} {\bibfnamefont {Javier}\ \bibnamefont
  {Redondo}}, \ and\ \bibinfo {author} {\bibfnamefont {Andreas}\ \bibnamefont
  {Ringwald}},\ }\bibfield  {title} {\enquote {\bibinfo {title} {{Naturally
  Light Hidden Photons in LARGE Volume String Compactifications}},}\ }\href
  {\doibase 10.1088/1126-6708/2009/11/027} {\bibfield  {journal} {\bibinfo
  {journal} {JHEP}\ }\textbf {\bibinfo {volume} {11}},\ \bibinfo {pages} {027}
  (\bibinfo {year} {2009})},\ \Eprint {http://arxiv.org/abs/0909.0515}
  {arXiv:0909.0515 [hep-ph]} \BibitemShut {NoStop}%
\bibitem [{\citenamefont {Shiu}\ \emph {et~al.}(2013)\citenamefont {Shiu},
  \citenamefont {Soler},\ and\ \citenamefont {Ye}}]{Shiu:2013wxa}%
  \BibitemOpen
  \bibfield  {author} {\bibinfo {author} {\bibfnamefont {Gary}\ \bibnamefont
  {Shiu}}, \bibinfo {author} {\bibfnamefont {Pablo}\ \bibnamefont {Soler}}, \
  and\ \bibinfo {author} {\bibfnamefont {Fang}\ \bibnamefont {Ye}},\ }\bibfield
   {title} {\enquote {\bibinfo {title} {{Milli-Charged Dark Matter in Quantum
  Gravity and String Theory}},}\ }\href {\doibase
  10.1103/PhysRevLett.110.241304} {\bibfield  {journal} {\bibinfo  {journal}
  {Phys. Rev. Lett.}\ }\textbf {\bibinfo {volume} {110}},\ \bibinfo {pages}
  {241304} (\bibinfo {year} {2013})},\ \Eprint {http://arxiv.org/abs/1302.5471}
  {arXiv:1302.5471 [hep-th]} \BibitemShut {NoStop}%
\bibitem [{\citenamefont {An}\ \emph {et~al.}(2013{\natexlab{a}})\citenamefont
  {An}, \citenamefont {Pospelov},\ and\ \citenamefont {Pradler}}]{An:2013yua}%
  \BibitemOpen
  \bibfield  {author} {\bibinfo {author} {\bibfnamefont {Haipeng}\ \bibnamefont
  {An}}, \bibinfo {author} {\bibfnamefont {Maxim}\ \bibnamefont {Pospelov}}, \
  and\ \bibinfo {author} {\bibfnamefont {Josef}\ \bibnamefont {Pradler}},\
  }\bibfield  {title} {\enquote {\bibinfo {title} {{Dark Matter Detectors as
  Dark Photon Helioscopes}},}\ }\href {\doibase 10.1103/PhysRevLett.111.041302}
  {\bibfield  {journal} {\bibinfo  {journal} {Phys. Rev. Lett.}\ }\textbf
  {\bibinfo {volume} {111}},\ \bibinfo {pages} {041302} (\bibinfo {year}
  {2013}{\natexlab{a}})},\ \Eprint {http://arxiv.org/abs/1304.3461}
  {arXiv:1304.3461 [hep-ph]} \BibitemShut {NoStop}%
\bibitem [{\citenamefont {An}\ \emph {et~al.}(2013{\natexlab{b}})\citenamefont
  {An}, \citenamefont {Pospelov},\ and\ \citenamefont {Pradler}}]{An:2013yfc}%
  \BibitemOpen
  \bibfield  {author} {\bibinfo {author} {\bibfnamefont {Haipeng}\ \bibnamefont
  {An}}, \bibinfo {author} {\bibfnamefont {Maxim}\ \bibnamefont {Pospelov}}, \
  and\ \bibinfo {author} {\bibfnamefont {Josef}\ \bibnamefont {Pradler}},\
  }\bibfield  {title} {\enquote {\bibinfo {title} {{New stellar constraints on
  dark photons}},}\ }\href {\doibase 10.1016/j.physletb.2013.07.008} {\bibfield
   {journal} {\bibinfo  {journal} {Phys. Lett. B}\ }\textbf {\bibinfo {volume}
  {725}},\ \bibinfo {pages} {190--195} (\bibinfo {year}
  {2013}{\natexlab{b}})},\ \Eprint {http://arxiv.org/abs/1302.3884}
  {arXiv:1302.3884 [hep-ph]} \BibitemShut {NoStop}%
\bibitem [{\citenamefont {Redondo}\ and\ \citenamefont
  {Raffelt}(2013)}]{Redondo:2013lna}%
  \BibitemOpen
  \bibfield  {author} {\bibinfo {author} {\bibfnamefont {Javier}\ \bibnamefont
  {Redondo}}\ and\ \bibinfo {author} {\bibfnamefont {Georg}\ \bibnamefont
  {Raffelt}},\ }\bibfield  {title} {\enquote {\bibinfo {title} {{Solar
  constraints on hidden photons re-visited}},}\ }\href {\doibase
  10.1088/1475-7516/2013/08/034} {\bibfield  {journal} {\bibinfo  {journal}
  {JCAP}\ }\textbf {\bibinfo {volume} {08}},\ \bibinfo {pages} {034} (\bibinfo
  {year} {2013})},\ \Eprint {http://arxiv.org/abs/1305.2920} {arXiv:1305.2920
  [hep-ph]} \BibitemShut {NoStop}%
\bibitem [{\citenamefont {Giannotti}\ \emph {et~al.}(2016)\citenamefont
  {Giannotti}, \citenamefont {Irastorza}, \citenamefont {Redondo},\ and\
  \citenamefont {Ringwald}}]{Giannotti:2015kwo}%
  \BibitemOpen
  \bibfield  {author} {\bibinfo {author} {\bibfnamefont {Maurizio}\
  \bibnamefont {Giannotti}}, \bibinfo {author} {\bibfnamefont {Igor}\
  \bibnamefont {Irastorza}}, \bibinfo {author} {\bibfnamefont {Javier}\
  \bibnamefont {Redondo}}, \ and\ \bibinfo {author} {\bibfnamefont {Andreas}\
  \bibnamefont {Ringwald}},\ }\bibfield  {title} {\enquote {\bibinfo {title}
  {{Cool WISPs for stellar cooling excesses}},}\ }\href {\doibase
  10.1088/1475-7516/2016/05/057} {\bibfield  {journal} {\bibinfo  {journal}
  {JCAP}\ }\textbf {\bibinfo {volume} {05}},\ \bibinfo {pages} {057} (\bibinfo
  {year} {2016})},\ \Eprint {http://arxiv.org/abs/1512.08108} {arXiv:1512.08108
  [astro-ph.HE]} \BibitemShut {NoStop}%
\bibitem [{\citenamefont {Chang}\ \emph {et~al.}(2018)\citenamefont {Chang},
  \citenamefont {Essig},\ and\ \citenamefont {McDermott}}]{Chang:2018rso}%
  \BibitemOpen
  \bibfield  {author} {\bibinfo {author} {\bibfnamefont {Jae~Hyeok}\
  \bibnamefont {Chang}}, \bibinfo {author} {\bibfnamefont {Rouven}\
  \bibnamefont {Essig}}, \ and\ \bibinfo {author} {\bibfnamefont {Samuel~D.}\
  \bibnamefont {McDermott}},\ }\bibfield  {title} {\enquote {\bibinfo {title}
  {{Supernova 1987A Constraints on Sub-GeV Dark Sectors, Millicharged
  Particles, the QCD Axion, and an Axion-like Particle}},}\ }\href {\doibase
  10.1007/JHEP09(2018)051} {\bibfield  {journal} {\bibinfo  {journal} {JHEP}\
  }\textbf {\bibinfo {volume} {09}},\ \bibinfo {pages} {051} (\bibinfo {year}
  {2018})},\ \Eprint {http://arxiv.org/abs/1803.00993} {arXiv:1803.00993
  [hep-ph]} \BibitemShut {NoStop}%
\bibitem [{\citenamefont {Lasenby}\ and\ \citenamefont
  {Van~Tilburg}(2021)}]{Lasenby:2020goo}%
  \BibitemOpen
  \bibfield  {author} {\bibinfo {author} {\bibfnamefont {Robert}\ \bibnamefont
  {Lasenby}}\ and\ \bibinfo {author} {\bibfnamefont {Ken}\ \bibnamefont
  {Van~Tilburg}},\ }\bibfield  {title} {\enquote {\bibinfo {title} {{Dark
  photons in the solar basin}},}\ }\href {\doibase 10.1103/PhysRevD.104.023020}
  {\bibfield  {journal} {\bibinfo  {journal} {Phys. Rev. D}\ }\textbf {\bibinfo
  {volume} {104}},\ \bibinfo {pages} {023020} (\bibinfo {year} {2021})},\
  \Eprint {http://arxiv.org/abs/2008.08594} {arXiv:2008.08594 [hep-ph]}
  \BibitemShut {NoStop}%
\bibitem [{\citenamefont {Dolan}\ \emph {et~al.}(2024)\citenamefont {Dolan},
  \citenamefont {Hiskens},\ and\ \citenamefont {Volkas}}]{Dolan:2023cjs}%
  \BibitemOpen
  \bibfield  {author} {\bibinfo {author} {\bibfnamefont {Matthew~J.}\
  \bibnamefont {Dolan}}, \bibinfo {author} {\bibfnamefont {Frederick~J.}\
  \bibnamefont {Hiskens}}, \ and\ \bibinfo {author} {\bibfnamefont
  {Raymond~R.}\ \bibnamefont {Volkas}},\ }\bibfield  {title} {\enquote
  {\bibinfo {title} {{Constraining dark photons with self-consistent
  simulations of globular cluster stars}},}\ }\href {\doibase
  10.1088/1475-7516/2024/05/099} {\bibfield  {journal} {\bibinfo  {journal}
  {JCAP}\ }\textbf {\bibinfo {volume} {05}},\ \bibinfo {pages} {099} (\bibinfo
  {year} {2024})},\ \Eprint {http://arxiv.org/abs/2306.13335} {arXiv:2306.13335
  [hep-ph]} \BibitemShut {NoStop}%
\bibitem [{\citenamefont {Fung}\ \emph {et~al.}(2024)\citenamefont {Fung},
  \citenamefont {Heeba}, \citenamefont {Liu}, \citenamefont {Muralidharan},
  \citenamefont {Schutz},\ and\ \citenamefont {Vincent}}]{Fung:2023euv}%
  \BibitemOpen
  \bibfield  {author} {\bibinfo {author} {\bibfnamefont {Audrey}\ \bibnamefont
  {Fung}}, \bibinfo {author} {\bibfnamefont {Saniya}\ \bibnamefont {Heeba}},
  \bibinfo {author} {\bibfnamefont {Qinrui}\ \bibnamefont {Liu}}, \bibinfo
  {author} {\bibfnamefont {Varun}\ \bibnamefont {Muralidharan}}, \bibinfo
  {author} {\bibfnamefont {Katelin}\ \bibnamefont {Schutz}}, \ and\ \bibinfo
  {author} {\bibfnamefont {Aaron~C.}\ \bibnamefont {Vincent}},\ }\bibfield
  {title} {\enquote {\bibinfo {title} {{New bounds on light millicharged
  particles from the tip of the red-giant branch}},}\ }\href {\doibase
  10.1103/PhysRevD.109.083011} {\bibfield  {journal} {\bibinfo  {journal}
  {Phys. Rev. D}\ }\textbf {\bibinfo {volume} {109}},\ \bibinfo {pages}
  {083011} (\bibinfo {year} {2024})},\ \Eprint
  {http://arxiv.org/abs/2309.06465} {arXiv:2309.06465 [hep-ph]} \BibitemShut
  {NoStop}%
\bibitem [{\citenamefont {Berlin}\ \emph {et~al.}(2019)\citenamefont {Berlin},
  \citenamefont {Blinov}, \citenamefont {Krnjaic}, \citenamefont {Schuster},\
  and\ \citenamefont {Toro}}]{Berlin:2018bsc}%
  \BibitemOpen
  \bibfield  {author} {\bibinfo {author} {\bibfnamefont {Asher}\ \bibnamefont
  {Berlin}}, \bibinfo {author} {\bibfnamefont {Nikita}\ \bibnamefont {Blinov}},
  \bibinfo {author} {\bibfnamefont {Gordan}\ \bibnamefont {Krnjaic}}, \bibinfo
  {author} {\bibfnamefont {Philip}\ \bibnamefont {Schuster}}, \ and\ \bibinfo
  {author} {\bibfnamefont {Natalia}\ \bibnamefont {Toro}},\ }\bibfield  {title}
  {\enquote {\bibinfo {title} {{Dark Matter, Millicharges, Axion and Scalar
  Particles, Gauge Bosons, and Other New Physics with LDMX}},}\ }\href
  {\doibase 10.1103/PhysRevD.99.075001} {\bibfield  {journal} {\bibinfo
  {journal} {Phys. Rev. D}\ }\textbf {\bibinfo {volume} {99}},\ \bibinfo
  {pages} {075001} (\bibinfo {year} {2019})},\ \Eprint
  {http://arxiv.org/abs/1807.01730} {arXiv:1807.01730 [hep-ph]} \BibitemShut
  {NoStop}%
\bibitem [{\citenamefont {\r{A}kesson}\ \emph {et~al.}(2018)\citenamefont
  {\r{A}kesson} \emph {et~al.}}]{LDMX:2018cma}%
  \BibitemOpen
  \bibfield  {author} {\bibinfo {author} {\bibfnamefont {Torsten}\ \bibnamefont
  {\r{A}kesson}} \emph {et~al.} (\bibinfo {collaboration} {LDMX}),\ }\bibfield
  {title} {\enquote {\bibinfo {title} {{Light Dark Matter eXperiment
  (LDMX)}},}\ }\href@noop {} {\  (\bibinfo {year} {2018})},\ \Eprint
  {http://arxiv.org/abs/1808.05219} {arXiv:1808.05219 [hep-ex]} \BibitemShut
  {NoStop}%
\bibitem [{\citenamefont {Ilten}\ \emph {et~al.}(2018)\citenamefont {Ilten},
  \citenamefont {Soreq}, \citenamefont {Williams},\ and\ \citenamefont
  {Xue}}]{Ilten:2018crw}%
  \BibitemOpen
  \bibfield  {author} {\bibinfo {author} {\bibfnamefont {Philip}\ \bibnamefont
  {Ilten}}, \bibinfo {author} {\bibfnamefont {Yotam}\ \bibnamefont {Soreq}},
  \bibinfo {author} {\bibfnamefont {Mike}\ \bibnamefont {Williams}}, \ and\
  \bibinfo {author} {\bibfnamefont {Wei}\ \bibnamefont {Xue}},\ }\bibfield
  {title} {\enquote {\bibinfo {title} {{Serendipity in dark photon
  searches}},}\ }\href {\doibase 10.1007/JHEP06(2018)004} {\bibfield  {journal}
  {\bibinfo  {journal} {JHEP}\ }\textbf {\bibinfo {volume} {06}},\ \bibinfo
  {pages} {004} (\bibinfo {year} {2018})},\ \Eprint
  {http://arxiv.org/abs/1801.04847} {arXiv:1801.04847 [hep-ph]} \BibitemShut
  {NoStop}%
\bibitem [{\citenamefont {Alimena}\ \emph {et~al.}(2020)\citenamefont {Alimena}
  \emph {et~al.}}]{Alimena:2019zri}%
  \BibitemOpen
  \bibfield  {author} {\bibinfo {author} {\bibfnamefont {Juliette}\
  \bibnamefont {Alimena}} \emph {et~al.},\ }\bibfield  {title} {\enquote
  {\bibinfo {title} {{Searching for long-lived particles beyond the Standard
  Model at the Large Hadron Collider}},}\ }\href {\doibase
  10.1088/1361-6471/ab4574} {\bibfield  {journal} {\bibinfo  {journal} {J.
  Phys. G}\ }\textbf {\bibinfo {volume} {47}},\ \bibinfo {pages} {090501}
  (\bibinfo {year} {2020})},\ \Eprint {http://arxiv.org/abs/1903.04497}
  {arXiv:1903.04497 [hep-ex]} \BibitemShut {NoStop}%
\bibitem [{\citenamefont {et~al. (Planck)}(2020)}]{2020}%
  \BibitemOpen
  \bibfield  {author} {\bibinfo {author} {\bibfnamefont {N.~Aghanim}\
  \bibnamefont {et~al. (Planck)}},\ }\bibfield  {title} {\enquote {\bibinfo
  {title} {Planck2018 results: Vi. cosmological parameters},}\ }\href {\doibase
  10.1051/0004-6361/201833910} {\bibfield  {journal} {\bibinfo  {journal}
  {Astronomy \&; Astrophysics}\ }\textbf {\bibinfo {volume} {641}},\ \bibinfo
  {pages} {A6} (\bibinfo {year} {2020})}\BibitemShut {NoStop}%
\bibitem [{\citenamefont {Adari}\ \emph {et~al.}(2023)\citenamefont {Adari}
  \emph {et~al.}}]{SENSEI:2023zdf}%
  \BibitemOpen
  \bibfield  {author} {\bibinfo {author} {\bibfnamefont {Prakruth}\
  \bibnamefont {Adari}} \emph {et~al.} (\bibinfo {collaboration} {SENSEI}),\
  }\bibfield  {title} {\enquote {\bibinfo {title} {{SENSEI: First
  Direct-Detection Results on sub-GeV Dark Matter from SENSEI at SNOLAB}},}\
  }\href@noop {} {\  (\bibinfo {year} {2023})},\ \Eprint
  {http://arxiv.org/abs/2312.13342} {arXiv:2312.13342 [astro-ph.CO]}
  \BibitemShut {NoStop}%
\bibitem [{\citenamefont {Arnquist}\ \emph {et~al.}(2023)\citenamefont
  {Arnquist} \emph {et~al.}}]{DAMIC-M:2023gxo}%
  \BibitemOpen
  \bibfield  {author} {\bibinfo {author} {\bibfnamefont {I.}~\bibnamefont
  {Arnquist}} \emph {et~al.} (\bibinfo {collaboration} {DAMIC-M}),\ }\bibfield
  {title} {\enquote {\bibinfo {title} {{First Constraints from DAMIC-M on
  Sub-GeV Dark-Matter Particles Interacting with Electrons}},}\ }\href
  {\doibase 10.1103/PhysRevLett.130.171003} {\bibfield  {journal} {\bibinfo
  {journal} {Phys. Rev. Lett.}\ }\textbf {\bibinfo {volume} {130}},\ \bibinfo
  {pages} {171003} (\bibinfo {year} {2023})},\ \Eprint
  {http://arxiv.org/abs/2302.02372} {arXiv:2302.02372 [hep-ex]} \BibitemShut
  {NoStop}%
\bibitem [{\citenamefont {Aguilar-Arevalo}\ \emph {et~al.}(2022)\citenamefont
  {Aguilar-Arevalo} \emph {et~al.}}]{Oscura:2022vmi}%
  \BibitemOpen
  \bibfield  {author} {\bibinfo {author} {\bibfnamefont {Alexis}\ \bibnamefont
  {Aguilar-Arevalo}} \emph {et~al.} (\bibinfo {collaboration} {Oscura}),\
  }\bibfield  {title} {\enquote {\bibinfo {title} {{The Oscura Experiment}},}\
  }\href@noop {} {\  (\bibinfo {year} {2022})},\ \Eprint
  {http://arxiv.org/abs/2202.10518} {arXiv:2202.10518 [astro-ph.IM]}
  \BibitemShut {NoStop}%
\bibitem [{\citenamefont {Knapen}\ \emph {et~al.}(2017)\citenamefont {Knapen},
  \citenamefont {Lin},\ and\ \citenamefont {Zurek}}]{Knapen:2017xzo}%
  \BibitemOpen
  \bibfield  {author} {\bibinfo {author} {\bibfnamefont {Simon}\ \bibnamefont
  {Knapen}}, \bibinfo {author} {\bibfnamefont {Tongyan}\ \bibnamefont {Lin}}, \
  and\ \bibinfo {author} {\bibfnamefont {Kathryn~M.}\ \bibnamefont {Zurek}},\
  }\bibfield  {title} {\enquote {\bibinfo {title} {{Light Dark Matter: Models
  and Constraints}},}\ }\href {\doibase 10.1103/PhysRevD.96.115021} {\bibfield
  {journal} {\bibinfo  {journal} {Phys. Rev. D}\ }\textbf {\bibinfo {volume}
  {96}},\ \bibinfo {pages} {115021} (\bibinfo {year} {2017})},\ \Eprint
  {http://arxiv.org/abs/1709.07882} {arXiv:1709.07882 [hep-ph]} \BibitemShut
  {NoStop}%
\bibitem [{\citenamefont {Berlin}\ \emph {et~al.}(2020)\citenamefont {Berlin},
  \citenamefont {D'Agnolo}, \citenamefont {Ellis}, \citenamefont {Schuster},\
  and\ \citenamefont {Toro}}]{Berlin_2020}%
  \BibitemOpen
  \bibfield  {author} {\bibinfo {author} {\bibfnamefont {Asher}\ \bibnamefont
  {Berlin}}, \bibinfo {author} {\bibfnamefont {Raffaele~Tito}\ \bibnamefont
  {D'Agnolo}}, \bibinfo {author} {\bibfnamefont {Sebastian A.~R.}\ \bibnamefont
  {Ellis}}, \bibinfo {author} {\bibfnamefont {Philip}\ \bibnamefont
  {Schuster}}, \ and\ \bibinfo {author} {\bibfnamefont {Natalia}\ \bibnamefont
  {Toro}},\ }\bibfield  {title} {\enquote {\bibinfo {title} {{Directly
  Deflecting Particle Dark Matter}},}\ }\href {\doibase
  10.1103/PhysRevLett.124.011801} {\bibfield  {journal} {\bibinfo  {journal}
  {Phys. Rev. Lett.}\ }\textbf {\bibinfo {volume} {124}},\ \bibinfo {pages}
  {011801} (\bibinfo {year} {2020})},\ \Eprint
  {http://arxiv.org/abs/1908.06982} {arXiv:1908.06982 [hep-ph]} \BibitemShut
  {NoStop}%
\bibitem [{\citenamefont {Knapen}\ \emph {et~al.}(2018)\citenamefont {Knapen},
  \citenamefont {Lin}, \citenamefont {Pyle},\ and\ \citenamefont
  {Zurek}}]{Knapen:2017ekk}%
  \BibitemOpen
  \bibfield  {author} {\bibinfo {author} {\bibfnamefont {Simon}\ \bibnamefont
  {Knapen}}, \bibinfo {author} {\bibfnamefont {Tongyan}\ \bibnamefont {Lin}},
  \bibinfo {author} {\bibfnamefont {Matt}\ \bibnamefont {Pyle}}, \ and\
  \bibinfo {author} {\bibfnamefont {Kathryn~M.}\ \bibnamefont {Zurek}},\
  }\bibfield  {title} {\enquote {\bibinfo {title} {{Detection of Light Dark
  Matter With Optical Phonons in Polar Materials}},}\ }\href {\doibase
  10.1016/j.physletb.2018.08.064} {\bibfield  {journal} {\bibinfo  {journal}
  {Phys. Lett. B}\ }\textbf {\bibinfo {volume} {785}},\ \bibinfo {pages}
  {386--390} (\bibinfo {year} {2018})},\ \Eprint
  {http://arxiv.org/abs/1712.06598} {arXiv:1712.06598 [hep-ph]} \BibitemShut
  {NoStop}%
\bibitem [{\citenamefont {Griffin}\ \emph {et~al.}(2018)\citenamefont
  {Griffin}, \citenamefont {Knapen}, \citenamefont {Lin},\ and\ \citenamefont
  {Zurek}}]{polarmaterials:2018bjn}%
  \BibitemOpen
  \bibfield  {author} {\bibinfo {author} {\bibfnamefont {Sinead}\ \bibnamefont
  {Griffin}}, \bibinfo {author} {\bibfnamefont {Simon}\ \bibnamefont {Knapen}},
  \bibinfo {author} {\bibfnamefont {Tongyan}\ \bibnamefont {Lin}}, \ and\
  \bibinfo {author} {\bibfnamefont {Kathryn~M.}\ \bibnamefont {Zurek}},\
  }\bibfield  {title} {\enquote {\bibinfo {title} {{Directional Detection of
  Light Dark Matter with Polar Materials}},}\ }\href {\doibase
  10.1103/PhysRevD.98.115034} {\bibfield  {journal} {\bibinfo  {journal} {Phys.
  Rev. D}\ }\textbf {\bibinfo {volume} {98}},\ \bibinfo {pages} {115034}
  (\bibinfo {year} {2018})},\ \Eprint {http://arxiv.org/abs/1807.10291}
  {arXiv:1807.10291 [hep-ph]} \BibitemShut {NoStop}%
\bibitem [{\citenamefont {Campbell-Deem}\ \emph {et~al.}(2022)\citenamefont
  {Campbell-Deem}, \citenamefont {Knapen}, \citenamefont {Lin},\ and\
  \citenamefont {Villarama}}]{Campbell-Deem:2022fqm}%
  \BibitemOpen
  \bibfield  {author} {\bibinfo {author} {\bibfnamefont {Brian}\ \bibnamefont
  {Campbell-Deem}}, \bibinfo {author} {\bibfnamefont {Simon}\ \bibnamefont
  {Knapen}}, \bibinfo {author} {\bibfnamefont {Tongyan}\ \bibnamefont {Lin}}, \
  and\ \bibinfo {author} {\bibfnamefont {Ethan}\ \bibnamefont {Villarama}},\
  }\bibfield  {title} {\enquote {\bibinfo {title} {{Dark matter direct
  detection from the single phonon to the nuclear recoil regime}},}\ }\href
  {\doibase 10.1103/PhysRevD.106.036019} {\bibfield  {journal} {\bibinfo
  {journal} {Phys. Rev. D}\ }\textbf {\bibinfo {volume} {106}},\ \bibinfo
  {pages} {036019} (\bibinfo {year} {2022})},\ \Eprint
  {http://arxiv.org/abs/2205.02250} {arXiv:2205.02250 [hep-ph]} \BibitemShut
  {NoStop}%
\bibitem [{\citenamefont {Tongyan~Lin}()}]{priv}%
  \BibitemOpen
  \bibfield  {author} {\bibinfo {author} {\bibfnamefont {Connor~Stratman}\
  \bibnamefont {Tongyan~Lin}},\ }\href@noop {} {}\bibinfo {howpublished}
  {personal communication}\BibitemShut {NoStop}%
\bibitem [{\citenamefont {Derenzo}\ \emph {et~al.}(2021)\citenamefont
  {Derenzo}, \citenamefont {Guo}, \citenamefont {Hertel}, \citenamefont
  {Sorensen}, \citenamefont {Suzuki},\ and\ \citenamefont
  {Zurek}}]{derenzosnowmass2021}%
  \BibitemOpen
  \bibfield  {author} {\bibinfo {author} {\bibfnamefont {S}~\bibnamefont
  {Derenzo}}, \bibinfo {author} {\bibfnamefont {W}~\bibnamefont {Guo}},
  \bibinfo {author} {\bibfnamefont {S}~\bibnamefont {Hertel}}, \bibinfo
  {author} {\bibfnamefont {P}~\bibnamefont {Sorensen}}, \bibinfo {author}
  {\bibfnamefont {A}~\bibnamefont {Suzuki}}, \ and\ \bibinfo {author}
  {\bibfnamefont {K}~\bibnamefont {Zurek}},\ }\bibfield  {title} {\enquote
  {\bibinfo {title} {Snowmass2021-letter of interest the tesseract dark matter
  project},}\ }\href@noop {} {\  (\bibinfo {year} {2021})}\BibitemShut
  {NoStop}%
\bibitem [{\citenamefont {Hochberg}\ \emph {et~al.}(2018)\citenamefont
  {Hochberg}, \citenamefont {Kahn}, \citenamefont {Lisanti}, \citenamefont
  {Zurek}, \citenamefont {Grushin}, \citenamefont {Ilan}, \citenamefont
  {Griffin}, \citenamefont {Liu}, \citenamefont {Weber},\ and\ \citenamefont
  {Neaton}}]{Hochberg:2017wce}%
  \BibitemOpen
  \bibfield  {author} {\bibinfo {author} {\bibfnamefont {Yonit}\ \bibnamefont
  {Hochberg}}, \bibinfo {author} {\bibfnamefont {Yonatan}\ \bibnamefont
  {Kahn}}, \bibinfo {author} {\bibfnamefont {Mariangela}\ \bibnamefont
  {Lisanti}}, \bibinfo {author} {\bibfnamefont {Kathryn~M.}\ \bibnamefont
  {Zurek}}, \bibinfo {author} {\bibfnamefont {Adolfo~G.}\ \bibnamefont
  {Grushin}}, \bibinfo {author} {\bibfnamefont {Roni}\ \bibnamefont {Ilan}},
  \bibinfo {author} {\bibfnamefont {Sin\'ead~M.}\ \bibnamefont {Griffin}},
  \bibinfo {author} {\bibfnamefont {Zhen-Fei}\ \bibnamefont {Liu}}, \bibinfo
  {author} {\bibfnamefont {Sophie~F.}\ \bibnamefont {Weber}}, \ and\ \bibinfo
  {author} {\bibfnamefont {Jeffrey~B.}\ \bibnamefont {Neaton}},\ }\bibfield
  {title} {\enquote {\bibinfo {title} {{Detection of sub-MeV Dark Matter with
  Three-Dimensional Dirac Materials}},}\ }\href {\doibase
  10.1103/PhysRevD.97.015004} {\bibfield  {journal} {\bibinfo  {journal} {Phys.
  Rev. D}\ }\textbf {\bibinfo {volume} {97}},\ \bibinfo {pages} {015004}
  (\bibinfo {year} {2018})},\ \Eprint {http://arxiv.org/abs/1708.08929}
  {arXiv:1708.08929 [hep-ph]} \BibitemShut {NoStop}%
\bibitem [{\citenamefont {Geilhufe}\ \emph {et~al.}(2020)\citenamefont
  {Geilhufe}, \citenamefont {Kahlhoefer},\ and\ \citenamefont
  {Winkler}}]{Geilhufe_2020DiracMaterials}%
  \BibitemOpen
  \bibfield  {author} {\bibinfo {author} {\bibfnamefont {R.~Matthias}\
  \bibnamefont {Geilhufe}}, \bibinfo {author} {\bibfnamefont {Felix}\
  \bibnamefont {Kahlhoefer}}, \ and\ \bibinfo {author} {\bibfnamefont
  {Martin~Wolfgang}\ \bibnamefont {Winkler}},\ }\bibfield  {title} {\enquote
  {\bibinfo {title} {Dirac materials for sub-mev dark matter detection: New
  targets and improved formalism},}\ }\href {\doibase
  10.1103/physrevd.101.055005} {\bibfield  {journal} {\bibinfo  {journal}
  {Physical Review D}\ }\textbf {\bibinfo {volume} {101}} (\bibinfo {year}
  {2020}),\ 10.1103/physrevd.101.055005}\BibitemShut {NoStop}%
\bibitem [{\citenamefont {Du}\ \emph {et~al.}(2024)\citenamefont {Du},
  \citenamefont {Ega\~na Ugrinovic}, \citenamefont {Essig},\ and\ \citenamefont
  {Sholapurkar}}]{Du:2022dxf}%
  \BibitemOpen
  \bibfield  {author} {\bibinfo {author} {\bibfnamefont {Peizhi}\ \bibnamefont
  {Du}}, \bibinfo {author} {\bibfnamefont {Daniel}\ \bibnamefont {Ega\~na
  Ugrinovic}}, \bibinfo {author} {\bibfnamefont {Rouven}\ \bibnamefont
  {Essig}}, \ and\ \bibinfo {author} {\bibfnamefont {Mukul}\ \bibnamefont
  {Sholapurkar}},\ }\bibfield  {title} {\enquote {\bibinfo {title} {{Doped
  semiconductor devices for sub-MeV dark matter detection}},}\ }\href {\doibase
  10.1103/PhysRevD.109.055009} {\bibfield  {journal} {\bibinfo  {journal}
  {Phys. Rev. D}\ }\textbf {\bibinfo {volume} {109}},\ \bibinfo {pages}
  {055009} (\bibinfo {year} {2024})},\ \Eprint
  {http://arxiv.org/abs/2212.04504} {arXiv:2212.04504 [hep-ph]} \BibitemShut
  {NoStop}%
\bibitem [{\citenamefont {Hochberg}\ \emph {et~al.}(2021)\citenamefont
  {Hochberg}, \citenamefont {Kahn}, \citenamefont {Kurinsky}, \citenamefont
  {Lehmann}, \citenamefont {Yu},\ and\ \citenamefont
  {Berggren}}]{Hochberg:2021pkt}%
  \BibitemOpen
  \bibfield  {author} {\bibinfo {author} {\bibfnamefont {Yonit}\ \bibnamefont
  {Hochberg}}, \bibinfo {author} {\bibfnamefont {Yonatan}\ \bibnamefont
  {Kahn}}, \bibinfo {author} {\bibfnamefont {Noah}\ \bibnamefont {Kurinsky}},
  \bibinfo {author} {\bibfnamefont {Benjamin~V.}\ \bibnamefont {Lehmann}},
  \bibinfo {author} {\bibfnamefont {To~Chin}\ \bibnamefont {Yu}}, \ and\
  \bibinfo {author} {\bibfnamefont {Karl~K.}\ \bibnamefont {Berggren}},\
  }\bibfield  {title} {\enquote {\bibinfo {title} {{Determining
  Dark-Matter\textendash{}Electron Scattering Rates from the Dielectric
  Function}},}\ }\href {\doibase 10.1103/PhysRevLett.127.151802} {\bibfield
  {journal} {\bibinfo  {journal} {Phys. Rev. Lett.}\ }\textbf {\bibinfo
  {volume} {127}},\ \bibinfo {pages} {151802} (\bibinfo {year} {2021})},\
  \Eprint {http://arxiv.org/abs/2101.08263} {arXiv:2101.08263 [hep-ph]}
  \BibitemShut {NoStop}%
\bibitem [{\citenamefont {Knapen}\ \emph {et~al.}(2021)\citenamefont {Knapen},
  \citenamefont {Kozaczuk},\ and\ \citenamefont {Lin}}]{Knapen:2021run}%
  \BibitemOpen
  \bibfield  {author} {\bibinfo {author} {\bibfnamefont {Simon}\ \bibnamefont
  {Knapen}}, \bibinfo {author} {\bibfnamefont {Jonathan}\ \bibnamefont
  {Kozaczuk}}, \ and\ \bibinfo {author} {\bibfnamefont {Tongyan}\ \bibnamefont
  {Lin}},\ }\bibfield  {title} {\enquote {\bibinfo {title} {{Dark
  matter-electron scattering in dielectrics}},}\ }\href {\doibase
  10.1103/PhysRevD.104.015031} {\bibfield  {journal} {\bibinfo  {journal}
  {Phys. Rev. D}\ }\textbf {\bibinfo {volume} {104}},\ \bibinfo {pages}
  {015031} (\bibinfo {year} {2021})},\ \Eprint
  {http://arxiv.org/abs/2101.08275} {arXiv:2101.08275 [hep-ph]} \BibitemShut
  {NoStop}%
\bibitem [{\citenamefont {Berlin}\ and\ \citenamefont
  {Schutz}(2022)}]{Berlin:2021kcm}%
  \BibitemOpen
  \bibfield  {author} {\bibinfo {author} {\bibfnamefont {Asher}\ \bibnamefont
  {Berlin}}\ and\ \bibinfo {author} {\bibfnamefont {Katelin}\ \bibnamefont
  {Schutz}},\ }\bibfield  {title} {\enquote {\bibinfo {title} {{Helioscope for
  gravitationally bound millicharged particles}},}\ }\href {\doibase
  10.1103/PhysRevD.105.095012} {\bibfield  {journal} {\bibinfo  {journal}
  {Phys. Rev. D}\ }\textbf {\bibinfo {volume} {105}},\ \bibinfo {pages}
  {095012} (\bibinfo {year} {2022})},\ \Eprint
  {http://arxiv.org/abs/2111.01796} {arXiv:2111.01796 [hep-ph]} \BibitemShut
  {NoStop}%
\bibitem [{\citenamefont {Hall}(2010)}]{JHEP}%
  \BibitemOpen
  \bibfield  {author} {\bibinfo {author} {\bibfnamefont {Jedamzik K.
  March-Russell J. et~al.}\ \bibnamefont {Hall}, \bibfnamefont {L.J.}},\
  }\bibfield  {title} {\enquote {\bibinfo {title} {{Freeze-in production of
  FIMP dark matter}},}\ }\href {\doibase 10.1007/JHEP03(2010)080} {\bibfield
  {journal} {\bibinfo  {journal} {J. High Energ. Phys.}\ }\textbf {\bibinfo
  {volume} {80}} (\bibinfo {year} {2010}),\
  10.1007/JHEP03(2010)080}\BibitemShut {NoStop}%
\bibitem [{\citenamefont {Chu}\ \emph {et~al.}(2012)\citenamefont {Chu},
  \citenamefont {Hambye},\ and\ \citenamefont {Tytgat}}]{Chu:2011be}%
  \BibitemOpen
  \bibfield  {author} {\bibinfo {author} {\bibfnamefont {Xiaoyong}\
  \bibnamefont {Chu}}, \bibinfo {author} {\bibfnamefont {Thomas}\ \bibnamefont
  {Hambye}}, \ and\ \bibinfo {author} {\bibfnamefont {Michel H.~G.}\
  \bibnamefont {Tytgat}},\ }\bibfield  {title} {\enquote {\bibinfo {title}
  {{The Four Basic Ways of Creating Dark Matter Through a Portal}},}\ }\href
  {\doibase 10.1088/1475-7516/2012/05/034} {\bibfield  {journal} {\bibinfo
  {journal} {JCAP}\ }\textbf {\bibinfo {volume} {05}},\ \bibinfo {pages} {034}
  (\bibinfo {year} {2012})},\ \Eprint {http://arxiv.org/abs/1112.0493}
  {arXiv:1112.0493 [hep-ph]} \BibitemShut {NoStop}%
\bibitem [{\citenamefont {Bernal}\ \emph {et~al.}(2017)\citenamefont {Bernal},
  \citenamefont {Heikinheimo}, \citenamefont {Tenkanen}, \citenamefont
  {Tuominen},\ and\ \citenamefont {Vaskonen}}]{FIMPdawn}%
  \BibitemOpen
  \bibfield  {author} {\bibinfo {author} {\bibfnamefont {Nicol\'{a}s}\
  \bibnamefont {Bernal}}, \bibinfo {author} {\bibfnamefont {Matti}\
  \bibnamefont {Heikinheimo}}, \bibinfo {author} {\bibfnamefont {Tommi}\
  \bibnamefont {Tenkanen}}, \bibinfo {author} {\bibfnamefont {Kimmo}\
  \bibnamefont {Tuominen}}, \ and\ \bibinfo {author} {\bibfnamefont {Ville}\
  \bibnamefont {Vaskonen}},\ }\bibfield  {title} {\enquote {\bibinfo {title}
  {The dawn of fimp dark matter: A review of models and constraints},}\ }\href
  {\doibase 10.1142/S0217751X1730023X} {\bibfield  {journal} {\bibinfo
  {journal} {International Journal of Modern Physics A}\ }\textbf {\bibinfo
  {volume} {32}},\ \bibinfo {pages} {1730023} (\bibinfo {year} {2017})},\
  \Eprint {http://arxiv.org/abs/https://doi.org/10.1142/S0217751X1730023X}
  {https://doi.org/10.1142/S0217751X1730023X} \BibitemShut {NoStop}%
\bibitem [{\citenamefont {Dvorkin}\ \emph {et~al.}(2019)\citenamefont
  {Dvorkin}, \citenamefont {Lin},\ and\ \citenamefont {Schutz}}]{Dvorkin_2019}%
  \BibitemOpen
  \bibfield  {author} {\bibinfo {author} {\bibfnamefont {Cora}\ \bibnamefont
  {Dvorkin}}, \bibinfo {author} {\bibfnamefont {Tongyan}\ \bibnamefont {Lin}},
  \ and\ \bibinfo {author} {\bibfnamefont {Katelin}\ \bibnamefont {Schutz}},\
  }\bibfield  {title} {\enquote {\bibinfo {title} {Making dark matter out of
  light: Freeze-in from plasma effects},}\ }\href {\doibase
  10.1103/physrevd.99.115009} {\bibfield  {journal} {\bibinfo  {journal}
  {Physical Review D}\ }\textbf {\bibinfo {volume} {99}} (\bibinfo {year}
  {2019}),\ 10.1103/physrevd.99.115009}\BibitemShut {NoStop}%
\bibitem [{\citenamefont {Gan}\ and\ \citenamefont {Tsai}(2023)}]{Gan:2023jbs}%
  \BibitemOpen
  \bibfield  {author} {\bibinfo {author} {\bibfnamefont {Xucheng}\ \bibnamefont
  {Gan}}\ and\ \bibinfo {author} {\bibfnamefont {Yu-Dai}\ \bibnamefont
  {Tsai}},\ }\bibfield  {title} {\enquote {\bibinfo {title} {{Cosmic
  Millicharge Background and Reheating Probes}},}\ }\href@noop {} {\  (\bibinfo
  {year} {2023})},\ \Eprint {http://arxiv.org/abs/2308.07951} {arXiv:2308.07951
  [hep-ph]} \BibitemShut {NoStop}%
\bibitem [{\citenamefont {Langhoff}\ \emph {et~al.}(2022)\citenamefont
  {Langhoff}, \citenamefont {Outmezguine},\ and\ \citenamefont
  {Rodd}}]{Langhoff_2022}%
  \BibitemOpen
  \bibfield  {author} {\bibinfo {author} {\bibfnamefont {Kevin}\ \bibnamefont
  {Langhoff}}, \bibinfo {author} {\bibfnamefont {Nadav~Joseph}\ \bibnamefont
  {Outmezguine}}, \ and\ \bibinfo {author} {\bibfnamefont {Nicholas~L.}\
  \bibnamefont {Rodd}},\ }\bibfield  {title} {\enquote {\bibinfo {title}
  {Irreducible axion background},}\ }\href {\doibase
  10.1103/physrevlett.129.241101} {\bibfield  {journal} {\bibinfo  {journal}
  {Physical Review Letters}\ }\textbf {\bibinfo {volume} {129}} (\bibinfo
  {year} {2022}),\ 10.1103/physrevlett.129.241101}\BibitemShut {NoStop}%
\bibitem [{\citenamefont {D'Eramo}\ \emph {et~al.}(2024)\citenamefont
  {D'Eramo}, \citenamefont {Tesi},\ and\ \citenamefont
  {Vaskonen}}]{DEramo:2024lsk}%
  \BibitemOpen
  \bibfield  {author} {\bibinfo {author} {\bibfnamefont {Francesco}\
  \bibnamefont {D'Eramo}}, \bibinfo {author} {\bibfnamefont {Andrea}\
  \bibnamefont {Tesi}}, \ and\ \bibinfo {author} {\bibfnamefont {Ville}\
  \bibnamefont {Vaskonen}},\ }\bibfield  {title} {\enquote {\bibinfo {title}
  {{Irreducible cosmological backgrounds of a real scalar with a broken
  symmetry}},}\ }\href@noop {} {\  (\bibinfo {year} {2024})},\ \Eprint
  {http://arxiv.org/abs/2407.19997} {arXiv:2407.19997 [hep-ph]} \BibitemShut
  {NoStop}%
\bibitem [{\citenamefont {Essig}\ \emph
  {et~al.}(2012{\natexlab{a}})\citenamefont {Essig}, \citenamefont {Mardon},\
  and\ \citenamefont {Volansky}}]{DDe+e-}%
  \BibitemOpen
  \bibfield  {author} {\bibinfo {author} {\bibfnamefont {Rouven}\ \bibnamefont
  {Essig}}, \bibinfo {author} {\bibfnamefont {Jeremy}\ \bibnamefont {Mardon}},
  \ and\ \bibinfo {author} {\bibfnamefont {Tomer}\ \bibnamefont {Volansky}},\
  }\bibfield  {title} {\enquote {\bibinfo {title} {Direct detection of sub-gev
  dark matter},}\ }\href {\doibase 10.1103/PhysRevD.85.076007} {\bibfield
  {journal} {\bibinfo  {journal} {Phys. Rev. D}\ }\textbf {\bibinfo {volume}
  {85}},\ \bibinfo {pages} {076007} (\bibinfo {year}
  {2012}{\natexlab{a}})}\BibitemShut {NoStop}%
\bibitem [{\citenamefont {Essig}\ \emph
  {et~al.}(2012{\natexlab{b}})\citenamefont {Essig}, \citenamefont
  {Manalaysay}, \citenamefont {Mardon}, \citenamefont {Sorensen},\ and\
  \citenamefont {Volansky}}]{DDxenon}%
  \BibitemOpen
  \bibfield  {author} {\bibinfo {author} {\bibfnamefont {Rouven}\ \bibnamefont
  {Essig}}, \bibinfo {author} {\bibfnamefont {Aaron}\ \bibnamefont
  {Manalaysay}}, \bibinfo {author} {\bibfnamefont {Jeremy}\ \bibnamefont
  {Mardon}}, \bibinfo {author} {\bibfnamefont {Peter}\ \bibnamefont
  {Sorensen}}, \ and\ \bibinfo {author} {\bibfnamefont {Tomer}\ \bibnamefont
  {Volansky}},\ }\bibfield  {title} {\enquote {\bibinfo {title} {First direct
  detection limits on sub-gev dark matter from xenon10},}\ }\href {\doibase
  10.1103/PhysRevLett.109.021301} {\bibfield  {journal} {\bibinfo  {journal}
  {Phys. Rev. Lett.}\ }\textbf {\bibinfo {volume} {109}},\ \bibinfo {pages}
  {021301} (\bibinfo {year} {2012}{\natexlab{b}})}\BibitemShut {NoStop}%
\bibitem [{\citenamefont {Berlin}\ \emph {et~al.}(2023)\citenamefont {Berlin},
  \citenamefont {Dror}, \citenamefont {Gan},\ and\ \citenamefont
  {Ruderman}}]{Berlin:2022hmt}%
  \BibitemOpen
  \bibfield  {author} {\bibinfo {author} {\bibfnamefont {Asher}\ \bibnamefont
  {Berlin}}, \bibinfo {author} {\bibfnamefont {Jeff~A.}\ \bibnamefont {Dror}},
  \bibinfo {author} {\bibfnamefont {Xucheng}\ \bibnamefont {Gan}}, \ and\
  \bibinfo {author} {\bibfnamefont {Joshua~T.}\ \bibnamefont {Ruderman}},\
  }\bibfield  {title} {\enquote {\bibinfo {title} {{Millicharged relics reveal
  massless dark photons}},}\ }\href {\doibase 10.1007/JHEP05(2023)046}
  {\bibfield  {journal} {\bibinfo  {journal} {JHEP}\ }\textbf {\bibinfo
  {volume} {05}},\ \bibinfo {pages} {046} (\bibinfo {year} {2023})},\ \Eprint
  {http://arxiv.org/abs/2211.05139} {arXiv:2211.05139 [hep-ph]} \BibitemShut
  {NoStop}%
\bibitem [{\citenamefont {Thomas}\ \emph {et~al.}(2020)\citenamefont {Thomas},
  \citenamefont {Dezen}, \citenamefont {Grohs},\ and\ \citenamefont
  {Kishimoto}}]{Thomas_2020}%
  \BibitemOpen
  \bibfield  {author} {\bibinfo {author} {\bibfnamefont {Luke~C.}\ \bibnamefont
  {Thomas}}, \bibinfo {author} {\bibfnamefont {Ted}\ \bibnamefont {Dezen}},
  \bibinfo {author} {\bibfnamefont {Evan~B.}\ \bibnamefont {Grohs}}, \ and\
  \bibinfo {author} {\bibfnamefont {Chad~T.}\ \bibnamefont {Kishimoto}},\
  }\bibfield  {title} {\enquote {\bibinfo {title} {Electron-positron
  annihilation freeze-out in the early universe},}\ }\href {\doibase
  10.1103/physrevd.101.063507} {\bibfield  {journal} {\bibinfo  {journal}
  {Physical Review D}\ }\textbf {\bibinfo {volume} {101}} (\bibinfo {year}
  {2020}),\ 10.1103/physrevd.101.063507}\BibitemShut {NoStop}%
\bibitem [{\citenamefont {Heeba}\ \emph {et~al.}(2023)\citenamefont {Heeba},
  \citenamefont {Lin},\ and\ \citenamefont {Schutz}}]{Heeba:2023bik}%
  \BibitemOpen
  \bibfield  {author} {\bibinfo {author} {\bibfnamefont {Saniya}\ \bibnamefont
  {Heeba}}, \bibinfo {author} {\bibfnamefont {Tongyan}\ \bibnamefont {Lin}}, \
  and\ \bibinfo {author} {\bibfnamefont {Katelin}\ \bibnamefont {Schutz}},\
  }\bibfield  {title} {\enquote {\bibinfo {title} {{Inelastic freeze-in}},}\
  }\href {\doibase 10.1103/PhysRevD.108.095016} {\bibfield  {journal} {\bibinfo
   {journal} {Phys. Rev. D}\ }\textbf {\bibinfo {volume} {108}},\ \bibinfo
  {pages} {095016} (\bibinfo {year} {2023})},\ \Eprint
  {http://arxiv.org/abs/2304.06072} {arXiv:2304.06072 [hep-ph]} \BibitemShut
  {NoStop}%
\bibitem [{\citenamefont {Evans}\ \emph {et~al.}(2018)\citenamefont {Evans},
  \citenamefont {Gori},\ and\ \citenamefont {Shelton}}]{Evans:2017kti}%
  \BibitemOpen
  \bibfield  {author} {\bibinfo {author} {\bibfnamefont {Jared~A.}\
  \bibnamefont {Evans}}, \bibinfo {author} {\bibfnamefont {Stefania}\
  \bibnamefont {Gori}}, \ and\ \bibinfo {author} {\bibfnamefont {Jessie}\
  \bibnamefont {Shelton}},\ }\bibfield  {title} {\enquote {\bibinfo {title}
  {{Looking for the WIMP Next Door}},}\ }\href {\doibase
  10.1007/JHEP02(2018)100} {\bibfield  {journal} {\bibinfo  {journal} {JHEP}\
  }\textbf {\bibinfo {volume} {02}},\ \bibinfo {pages} {100} (\bibinfo {year}
  {2018})},\ \Eprint {http://arxiv.org/abs/1712.03974} {arXiv:1712.03974
  [hep-ph]} \BibitemShut {NoStop}%
\bibitem [{\citenamefont {Comelli}\ and\ \citenamefont
  {Espinosa}(1997)}]{PhysRevD.55.6253}%
  \BibitemOpen
  \bibfield  {author} {\bibinfo {author} {\bibfnamefont {D.}~\bibnamefont
  {Comelli}}\ and\ \bibinfo {author} {\bibfnamefont {J.~R.}\ \bibnamefont
  {Espinosa}},\ }\bibfield  {title} {\enquote {\bibinfo {title} {Bosonic
  thermal masses in supersymmetry},}\ }\href {\doibase
  10.1103/PhysRevD.55.6253} {\bibfield  {journal} {\bibinfo  {journal} {Phys.
  Rev. D}\ }\textbf {\bibinfo {volume} {55}},\ \bibinfo {pages} {6253--6263}
  (\bibinfo {year} {1997})}\BibitemShut {NoStop}%
\bibitem [{\citenamefont {Heeba}\ and\ \citenamefont
  {Kahlhoefer}(2020)}]{Heeba:2019jho}%
  \BibitemOpen
  \bibfield  {author} {\bibinfo {author} {\bibfnamefont {Saniya}\ \bibnamefont
  {Heeba}}\ and\ \bibinfo {author} {\bibfnamefont {Felix}\ \bibnamefont
  {Kahlhoefer}},\ }\bibfield  {title} {\enquote {\bibinfo {title} {{Probing the
  freeze-in mechanism in dark matter models with U(1)' gauge extensions}},}\
  }\href {\doibase 10.1103/PhysRevD.101.035043} {\bibfield  {journal} {\bibinfo
   {journal} {Phys. Rev. D}\ }\textbf {\bibinfo {volume} {101}},\ \bibinfo
  {pages} {035043} (\bibinfo {year} {2020})},\ \Eprint
  {http://arxiv.org/abs/1908.09834} {arXiv:1908.09834 [hep-ph]} \BibitemShut
  {NoStop}%
\bibitem [{\citenamefont {Reece}(2019)}]{Reece:2018zvv}%
  \BibitemOpen
  \bibfield  {author} {\bibinfo {author} {\bibfnamefont {Matthew}\ \bibnamefont
  {Reece}},\ }\bibfield  {title} {\enquote {\bibinfo {title} {{Photon Masses in
  the Landscape and the Swampland}},}\ }\href {\doibase
  10.1007/JHEP07(2019)181} {\bibfield  {journal} {\bibinfo  {journal} {JHEP}\
  }\textbf {\bibinfo {volume} {07}},\ \bibinfo {pages} {181} (\bibinfo {year}
  {2019})},\ \Eprint {http://arxiv.org/abs/1808.09966} {arXiv:1808.09966
  [hep-th]} \BibitemShut {NoStop}%
\bibitem [{\citenamefont {Gherghetta}\ \emph {et~al.}(2019)\citenamefont
  {Gherghetta}, \citenamefont {Kersten}, \citenamefont {Olive},\ and\
  \citenamefont {Pospelov}}]{Gherghetta:2019coi}%
  \BibitemOpen
  \bibfield  {author} {\bibinfo {author} {\bibfnamefont {Tony}\ \bibnamefont
  {Gherghetta}}, \bibinfo {author} {\bibfnamefont {J\"orn}\ \bibnamefont
  {Kersten}}, \bibinfo {author} {\bibfnamefont {Keith}\ \bibnamefont {Olive}},
  \ and\ \bibinfo {author} {\bibfnamefont {Maxim}\ \bibnamefont {Pospelov}},\
  }\bibfield  {title} {\enquote {\bibinfo {title} {{Evaluating the price of
  tiny kinetic mixing}},}\ }\href {\doibase 10.1103/PhysRevD.100.095001}
  {\bibfield  {journal} {\bibinfo  {journal} {Phys. Rev. D}\ }\textbf {\bibinfo
  {volume} {100}},\ \bibinfo {pages} {095001} (\bibinfo {year} {2019})},\
  \Eprint {http://arxiv.org/abs/1909.00696} {arXiv:1909.00696 [hep-ph]}
  \BibitemShut {NoStop}%
\bibitem [{\citenamefont {Tulin}\ and\ \citenamefont
  {Yu}(2018)}]{Tulin:2017ara}%
  \BibitemOpen
  \bibfield  {author} {\bibinfo {author} {\bibfnamefont {Sean}\ \bibnamefont
  {Tulin}}\ and\ \bibinfo {author} {\bibfnamefont {Hai-Bo}\ \bibnamefont
  {Yu}},\ }\bibfield  {title} {\enquote {\bibinfo {title} {{Dark Matter
  Self-interactions and Small Scale Structure}},}\ }\href {\doibase
  10.1016/j.physrep.2017.11.004} {\bibfield  {journal} {\bibinfo  {journal}
  {Phys. Rept.}\ }\textbf {\bibinfo {volume} {730}},\ \bibinfo {pages} {1--57}
  (\bibinfo {year} {2018})},\ \Eprint {http://arxiv.org/abs/1705.02358}
  {arXiv:1705.02358 [hep-ph]} \BibitemShut {NoStop}%
\bibitem [{\citenamefont {Kummer}\ \emph {et~al.}(2019)\citenamefont {Kummer},
  \citenamefont {Br\"uggen}, \citenamefont {Dolag}, \citenamefont
  {Kahlhoefer},\ and\ \citenamefont {Schmidt-Hoberg}}]{Kummer:2019yrb}%
  \BibitemOpen
  \bibfield  {author} {\bibinfo {author} {\bibfnamefont {Janis}\ \bibnamefont
  {Kummer}}, \bibinfo {author} {\bibfnamefont {Marcus}\ \bibnamefont
  {Br\"uggen}}, \bibinfo {author} {\bibfnamefont {Klaus}\ \bibnamefont
  {Dolag}}, \bibinfo {author} {\bibfnamefont {Felix}\ \bibnamefont
  {Kahlhoefer}}, \ and\ \bibinfo {author} {\bibfnamefont {Kai}\ \bibnamefont
  {Schmidt-Hoberg}},\ }\bibfield  {title} {\enquote {\bibinfo {title}
  {{Simulations of core formation for frequent dark matter
  self-interactions}},}\ }\href {\doibase 10.1093/mnras/stz1261} {\bibfield
  {journal} {\bibinfo  {journal} {Mon. Not. Roy. Astron. Soc.}\ }\textbf
  {\bibinfo {volume} {487}},\ \bibinfo {pages} {354--363} (\bibinfo {year}
  {2019})},\ \Eprint {http://arxiv.org/abs/1902.02330} {arXiv:1902.02330
  [astro-ph.CO]} \BibitemShut {NoStop}%
\bibitem [{\citenamefont {Read}(2014)}]{Read:2014qva}%
  \BibitemOpen
  \bibfield  {author} {\bibinfo {author} {\bibfnamefont {J.~I.}\ \bibnamefont
  {Read}},\ }\bibfield  {title} {\enquote {\bibinfo {title} {{The Local Dark
  Matter Density}},}\ }\href {\doibase 10.1088/0954-3899/41/6/063101}
  {\bibfield  {journal} {\bibinfo  {journal} {J. Phys. G}\ }\textbf {\bibinfo
  {volume} {41}},\ \bibinfo {pages} {063101} (\bibinfo {year} {2014})},\
  \Eprint {http://arxiv.org/abs/1404.1938} {arXiv:1404.1938 [astro-ph.GA]}
  \BibitemShut {NoStop}%
\bibitem [{\citenamefont {de~Salas}\ and\ \citenamefont
  {Widmark}(2021)}]{deSalas:2020hbh}%
  \BibitemOpen
  \bibfield  {author} {\bibinfo {author} {\bibfnamefont {Pablo~F.}\
  \bibnamefont {de~Salas}}\ and\ \bibinfo {author} {\bibfnamefont {Axel}\
  \bibnamefont {Widmark}},\ }\bibfield  {title} {\enquote {\bibinfo {title}
  {{Dark matter local density determination: recent observations and future
  prospects}},}\ }\href {\doibase 10.1088/1361-6633/ac24e7} {\bibfield
  {journal} {\bibinfo  {journal} {Rept. Prog. Phys.}\ }\textbf {\bibinfo
  {volume} {84}},\ \bibinfo {pages} {104901} (\bibinfo {year} {2021})},\
  \Eprint {http://arxiv.org/abs/2012.11477} {arXiv:2012.11477 [astro-ph.GA]}
  \BibitemShut {NoStop}%
\bibitem [{\citenamefont {Hambye}\ \emph {et~al.}(2018)\citenamefont {Hambye},
  \citenamefont {Tytgat}, \citenamefont {Vandecasteele},\ and\ \citenamefont
  {Vanderheyden}}]{Hambye:2018dpi}%
  \BibitemOpen
  \bibfield  {author} {\bibinfo {author} {\bibfnamefont {Thomas}\ \bibnamefont
  {Hambye}}, \bibinfo {author} {\bibfnamefont {Michel H.~G.}\ \bibnamefont
  {Tytgat}}, \bibinfo {author} {\bibfnamefont {J\'er\^ome}\ \bibnamefont
  {Vandecasteele}}, \ and\ \bibinfo {author} {\bibfnamefont {Laurent}\
  \bibnamefont {Vanderheyden}},\ }\bibfield  {title} {\enquote {\bibinfo
  {title} {{Dark matter direct detection is testing freeze-in}},}\ }\href
  {\doibase 10.1103/PhysRevD.98.075017} {\bibfield  {journal} {\bibinfo
  {journal} {Phys. Rev. D}\ }\textbf {\bibinfo {volume} {98}},\ \bibinfo
  {pages} {075017} (\bibinfo {year} {2018})},\ \Eprint
  {http://arxiv.org/abs/1807.05022} {arXiv:1807.05022 [hep-ph]} \BibitemShut
  {NoStop}%
\bibitem [{\citenamefont {Lebedev}(2023)}]{Lebedev:2022cic}%
  \BibitemOpen
  \bibfield  {author} {\bibinfo {author} {\bibfnamefont {Oleg}\ \bibnamefont
  {Lebedev}},\ }\bibfield  {title} {\enquote {\bibinfo {title} {{Scalar
  overproduction in standard cosmology and predictivity of non-thermal dark
  matter}},}\ }\href {\doibase 10.1088/1475-7516/2023/02/032} {\bibfield
  {journal} {\bibinfo  {journal} {JCAP}\ }\textbf {\bibinfo {volume} {02}},\
  \bibinfo {pages} {032} (\bibinfo {year} {2023})},\ \Eprint
  {http://arxiv.org/abs/2210.02293} {arXiv:2210.02293 [hep-ph]} \BibitemShut
  {NoStop}%
\bibitem [{\citenamefont {Hasegawa}\ \emph {et~al.}(2019)\citenamefont
  {Hasegawa}, \citenamefont {Hiroshima}, \citenamefont {Kohri}, \citenamefont
  {Hansen}, \citenamefont {Tram},\ and\ \citenamefont
  {Hannestad}}]{Hasegawa:2019jsa}%
  \BibitemOpen
  \bibfield  {author} {\bibinfo {author} {\bibfnamefont {Takuya}\ \bibnamefont
  {Hasegawa}}, \bibinfo {author} {\bibfnamefont {Nagisa}\ \bibnamefont
  {Hiroshima}}, \bibinfo {author} {\bibfnamefont {Kazunori}\ \bibnamefont
  {Kohri}}, \bibinfo {author} {\bibfnamefont {Rasmus S.~L.}\ \bibnamefont
  {Hansen}}, \bibinfo {author} {\bibfnamefont {Thomas}\ \bibnamefont {Tram}}, \
  and\ \bibinfo {author} {\bibfnamefont {Steen}\ \bibnamefont {Hannestad}},\
  }\bibfield  {title} {\enquote {\bibinfo {title} {{MeV-scale reheating
  temperature and thermalization of oscillating neutrinos by radiative and
  hadronic decays of massive particles}},}\ }\href {\doibase
  10.1088/1475-7516/2019/12/012} {\bibfield  {journal} {\bibinfo  {journal}
  {JCAP}\ }\textbf {\bibinfo {volume} {12}},\ \bibinfo {pages} {012} (\bibinfo
  {year} {2019})},\ \Eprint {http://arxiv.org/abs/1908.10189} {arXiv:1908.10189
  [hep-ph]} \BibitemShut {NoStop}%
\bibitem [{\citenamefont {de~Salas}\ \emph {et~al.}(2015)\citenamefont
  {de~Salas}, \citenamefont {Lattanzi}, \citenamefont {Mangano}, \citenamefont
  {Miele}, \citenamefont {Pastor},\ and\ \citenamefont
  {Pisanti}}]{deSalas:2015glj}%
  \BibitemOpen
  \bibfield  {author} {\bibinfo {author} {\bibfnamefont {P.~F.}\ \bibnamefont
  {de~Salas}}, \bibinfo {author} {\bibfnamefont {M.}~\bibnamefont {Lattanzi}},
  \bibinfo {author} {\bibfnamefont {G.}~\bibnamefont {Mangano}}, \bibinfo
  {author} {\bibfnamefont {G.}~\bibnamefont {Miele}}, \bibinfo {author}
  {\bibfnamefont {S.}~\bibnamefont {Pastor}}, \ and\ \bibinfo {author}
  {\bibfnamefont {O.}~\bibnamefont {Pisanti}},\ }\bibfield  {title} {\enquote
  {\bibinfo {title} {{Bounds on very low reheating scenarios after Planck}},}\
  }\href {\doibase 10.1103/PhysRevD.92.123534} {\bibfield  {journal} {\bibinfo
  {journal} {Phys. Rev. D}\ }\textbf {\bibinfo {volume} {92}},\ \bibinfo
  {pages} {123534} (\bibinfo {year} {2015})},\ \Eprint
  {http://arxiv.org/abs/1511.00672} {arXiv:1511.00672 [astro-ph.CO]}
  \BibitemShut {NoStop}%
\bibitem [{\citenamefont {Hannestad}(2004)}]{Hannestad:2004px}%
  \BibitemOpen
  \bibfield  {author} {\bibinfo {author} {\bibfnamefont {Steen}\ \bibnamefont
  {Hannestad}},\ }\bibfield  {title} {\enquote {\bibinfo {title} {{What is the
  lowest possible reheating temperature?}}}\ }\href {\doibase
  10.1103/PhysRevD.70.043506} {\bibfield  {journal} {\bibinfo  {journal} {Phys.
  Rev. D}\ }\textbf {\bibinfo {volume} {70}},\ \bibinfo {pages} {043506}
  (\bibinfo {year} {2004})},\ \Eprint {http://arxiv.org/abs/astro-ph/0403291}
  {arXiv:astro-ph/0403291} \BibitemShut {NoStop}%
\bibitem [{\citenamefont {Vinyoles}\ and\ \citenamefont
  {Vogel}(2016)}]{Vinyoles:2015khy}%
  \BibitemOpen
  \bibfield  {author} {\bibinfo {author} {\bibfnamefont {N\'uria}\ \bibnamefont
  {Vinyoles}}\ and\ \bibinfo {author} {\bibfnamefont {Hendrik}\ \bibnamefont
  {Vogel}},\ }\bibfield  {title} {\enquote {\bibinfo {title} {{Minicharged
  Particles from the Sun: A Cutting-Edge Bound}},}\ }\href {\doibase
  10.1088/1475-7516/2016/03/002} {\bibfield  {journal} {\bibinfo  {journal}
  {JCAP}\ }\textbf {\bibinfo {volume} {03}},\ \bibinfo {pages} {002} (\bibinfo
  {year} {2016})},\ \Eprint {http://arxiv.org/abs/1511.01122} {arXiv:1511.01122
  [hep-ph]} \BibitemShut {NoStop}%
\bibitem [{\citenamefont {Dvorkin}\ \emph {et~al.}(2021)\citenamefont
  {Dvorkin}, \citenamefont {Lin},\ and\ \citenamefont
  {Schutz}}]{Dvorkin:2020xga}%
  \BibitemOpen
  \bibfield  {author} {\bibinfo {author} {\bibfnamefont {Cora}\ \bibnamefont
  {Dvorkin}}, \bibinfo {author} {\bibfnamefont {Tongyan}\ \bibnamefont {Lin}},
  \ and\ \bibinfo {author} {\bibfnamefont {Katelin}\ \bibnamefont {Schutz}},\
  }\bibfield  {title} {\enquote {\bibinfo {title} {{Cosmology of Sub-MeV Dark
  Matter Freeze-In}},}\ }\href {\doibase 10.1103/PhysRevLett.127.111301}
  {\bibfield  {journal} {\bibinfo  {journal} {Phys. Rev. Lett.}\ }\textbf
  {\bibinfo {volume} {127}},\ \bibinfo {pages} {111301} (\bibinfo {year}
  {2021})},\ \Eprint {http://arxiv.org/abs/2011.08186} {arXiv:2011.08186
  [astro-ph.CO]} \BibitemShut {NoStop}%
\bibitem [{\citenamefont {Cruz}\ and\ \citenamefont
  {McQuinn}(2023)}]{Cruz:2022otv}%
  \BibitemOpen
  \bibfield  {author} {\bibinfo {author} {\bibfnamefont {Akaxia}\ \bibnamefont
  {Cruz}}\ and\ \bibinfo {author} {\bibfnamefont {Matthew}\ \bibnamefont
  {McQuinn}},\ }\bibfield  {title} {\enquote {\bibinfo {title} {{Astrophysical
  plasma instabilities induced by long-range interacting dark matter}},}\
  }\href {\doibase 10.1088/1475-7516/2023/04/028} {\bibfield  {journal}
  {\bibinfo  {journal} {JCAP}\ }\textbf {\bibinfo {volume} {04}},\ \bibinfo
  {pages} {028} (\bibinfo {year} {2023})},\ \Eprint
  {http://arxiv.org/abs/2202.12464} {arXiv:2202.12464 [astro-ph.CO]}
  \BibitemShut {NoStop}%
\bibitem [{\citenamefont {Stebbins}\ and\ \citenamefont
  {Krnjaic}(2019)}]{Stebbins:2019xjr}%
  \BibitemOpen
  \bibfield  {author} {\bibinfo {author} {\bibfnamefont {Albert}\ \bibnamefont
  {Stebbins}}\ and\ \bibinfo {author} {\bibfnamefont {Gordan}\ \bibnamefont
  {Krnjaic}},\ }\bibfield  {title} {\enquote {\bibinfo {title} {{New Limits on
  Charged Dark Matter from Large-Scale Coherent Magnetic Fields}},}\ }\href
  {\doibase 10.1088/1475-7516/2019/12/003} {\bibfield  {journal} {\bibinfo
  {journal} {JCAP}\ }\textbf {\bibinfo {volume} {12}},\ \bibinfo {pages} {003}
  (\bibinfo {year} {2019})},\ \Eprint {http://arxiv.org/abs/1908.05275}
  {arXiv:1908.05275 [astro-ph.CO]} \BibitemShut {NoStop}%
\bibitem [{\citenamefont {Lasenby}(2020)}]{Lasenby:2020rlf}%
  \BibitemOpen
  \bibfield  {author} {\bibinfo {author} {\bibfnamefont {Robert}\ \bibnamefont
  {Lasenby}},\ }\bibfield  {title} {\enquote {\bibinfo {title} {{Long range
  dark matter self-interactions and plasma instabilities}},}\ }\href {\doibase
  10.1088/1475-7516/2020/11/034} {\bibfield  {journal} {\bibinfo  {journal}
  {JCAP}\ }\textbf {\bibinfo {volume} {11}},\ \bibinfo {pages} {034} (\bibinfo
  {year} {2020})},\ \Eprint {http://arxiv.org/abs/2007.00667} {arXiv:2007.00667
  [hep-ph]} \BibitemShut {NoStop}%
\bibitem [{\citenamefont {Chuzhoy}\ and\ \citenamefont
  {Kolb}(2009)}]{Chuzhoy:2008zy}%
  \BibitemOpen
  \bibfield  {author} {\bibinfo {author} {\bibfnamefont {Leonid}\ \bibnamefont
  {Chuzhoy}}\ and\ \bibinfo {author} {\bibfnamefont {Edward~W.}\ \bibnamefont
  {Kolb}},\ }\bibfield  {title} {\enquote {\bibinfo {title} {{Reopening the
  window on charged dark matter}},}\ }\href {\doibase
  10.1088/1475-7516/2009/07/014} {\bibfield  {journal} {\bibinfo  {journal}
  {JCAP}\ }\textbf {\bibinfo {volume} {07}},\ \bibinfo {pages} {014} (\bibinfo
  {year} {2009})},\ \Eprint {http://arxiv.org/abs/0809.0436} {arXiv:0809.0436
  [astro-ph]} \BibitemShut {NoStop}%
\bibitem [{\citenamefont {Foot}(2011)}]{Foot:2010yz}%
  \BibitemOpen
  \bibfield  {author} {\bibinfo {author} {\bibfnamefont {R.}~\bibnamefont
  {Foot}},\ }\bibfield  {title} {\enquote {\bibinfo {title} {{Do magnetic
  fields prevent mirror particles from entering the galactic disk?}}}\ }\href
  {\doibase 10.1016/j.physletb.2011.04.012} {\bibfield  {journal} {\bibinfo
  {journal} {Phys. Lett. B}\ }\textbf {\bibinfo {volume} {699}},\ \bibinfo
  {pages} {230--232} (\bibinfo {year} {2011})},\ \Eprint
  {http://arxiv.org/abs/1011.5078} {arXiv:1011.5078 [hep-ph]} \BibitemShut
  {NoStop}%
\bibitem [{\citenamefont {Dunsky}\ \emph {et~al.}(2019)\citenamefont {Dunsky},
  \citenamefont {Hall},\ and\ \citenamefont {Harigaya}}]{Dunsky:2018mqs}%
  \BibitemOpen
  \bibfield  {author} {\bibinfo {author} {\bibfnamefont {David}\ \bibnamefont
  {Dunsky}}, \bibinfo {author} {\bibfnamefont {Lawrence~J.}\ \bibnamefont
  {Hall}}, \ and\ \bibinfo {author} {\bibfnamefont {Keisuke}\ \bibnamefont
  {Harigaya}},\ }\bibfield  {title} {\enquote {\bibinfo {title} {{CHAMP Cosmic
  Rays}},}\ }\href {\doibase 10.1088/1475-7516/2019/07/015} {\bibfield
  {journal} {\bibinfo  {journal} {JCAP}\ }\textbf {\bibinfo {volume} {07}},\
  \bibinfo {pages} {015} (\bibinfo {year} {2019})},\ \Eprint
  {http://arxiv.org/abs/1812.11116} {arXiv:1812.11116 [astro-ph.HE]}
  \BibitemShut {NoStop}%
\bibitem [{\citenamefont {Li}\ and\ \citenamefont {Lin}(2020)}]{Li:2020wyl}%
  \BibitemOpen
  \bibfield  {author} {\bibinfo {author} {\bibfnamefont {Jung-Tsung}\
  \bibnamefont {Li}}\ and\ \bibinfo {author} {\bibfnamefont {Tongyan}\
  \bibnamefont {Lin}},\ }\bibfield  {title} {\enquote {\bibinfo {title}
  {{Dynamics of millicharged dark matter in supernova remnants}},}\ }\href
  {\doibase 10.1103/PhysRevD.101.103034} {\bibfield  {journal} {\bibinfo
  {journal} {Phys. Rev. D}\ }\textbf {\bibinfo {volume} {101}},\ \bibinfo
  {pages} {103034} (\bibinfo {year} {2020})},\ \Eprint
  {http://arxiv.org/abs/2002.04625} {arXiv:2002.04625 [astro-ph.CO]}
  \BibitemShut {NoStop}%
\bibitem [{\citenamefont {Berlin}\ \emph {et~al.}(2024)\citenamefont {Berlin},
  \citenamefont {Liu}, \citenamefont {Pospelov},\ and\ \citenamefont
  {Ramani}}]{Berlin:2023zpn}%
  \BibitemOpen
  \bibfield  {author} {\bibinfo {author} {\bibfnamefont {Asher}\ \bibnamefont
  {Berlin}}, \bibinfo {author} {\bibfnamefont {Hongwan}\ \bibnamefont {Liu}},
  \bibinfo {author} {\bibfnamefont {Maxim}\ \bibnamefont {Pospelov}}, \ and\
  \bibinfo {author} {\bibfnamefont {Harikrishnan}\ \bibnamefont {Ramani}},\
  }\bibfield  {title} {\enquote {\bibinfo {title} {{Terrestrial density of
  strongly-coupled relics}},}\ }\href {\doibase 10.1103/PhysRevD.109.075027}
  {\bibfield  {journal} {\bibinfo  {journal} {Phys. Rev. D}\ }\textbf {\bibinfo
  {volume} {109}},\ \bibinfo {pages} {075027} (\bibinfo {year} {2024})},\
  \Eprint {http://arxiv.org/abs/2302.06619} {arXiv:2302.06619 [hep-ph]}
  \BibitemShut {NoStop}%
\bibitem [{\citenamefont {Berlin}\ \emph {et~al.}(2018)\citenamefont {Berlin},
  \citenamefont {Hooper}, \citenamefont {Krnjaic},\ and\ \citenamefont
  {McDermott}}]{PhysRevLett.121.011102}%
  \BibitemOpen
  \bibfield  {author} {\bibinfo {author} {\bibfnamefont {Asher}\ \bibnamefont
  {Berlin}}, \bibinfo {author} {\bibfnamefont {Dan}\ \bibnamefont {Hooper}},
  \bibinfo {author} {\bibfnamefont {Gordan}\ \bibnamefont {Krnjaic}}, \ and\
  \bibinfo {author} {\bibfnamefont {Samuel~D.}\ \bibnamefont {McDermott}},\
  }\bibfield  {title} {\enquote {\bibinfo {title} {Severely constraining
  dark-matter interpretations of the 21-cm anomaly},}\ }\href {\doibase
  10.1103/PhysRevLett.121.011102} {\bibfield  {journal} {\bibinfo  {journal}
  {Phys. Rev. Lett.}\ }\textbf {\bibinfo {volume} {121}},\ \bibinfo {pages}
  {011102} (\bibinfo {year} {2018})}\BibitemShut {NoStop}%
\bibitem [{\citenamefont {Barkana}\ \emph {et~al.}(2023)\citenamefont
  {Barkana}, \citenamefont {Fialkov}, \citenamefont {Liu},\ and\ \citenamefont
  {Outmezguine}}]{Barkana:2022hko}%
  \BibitemOpen
  \bibfield  {author} {\bibinfo {author} {\bibfnamefont {Rennan}\ \bibnamefont
  {Barkana}}, \bibinfo {author} {\bibfnamefont {Anastasia}\ \bibnamefont
  {Fialkov}}, \bibinfo {author} {\bibfnamefont {Hongwan}\ \bibnamefont {Liu}},
  \ and\ \bibinfo {author} {\bibfnamefont {Nadav~Joseph}\ \bibnamefont
  {Outmezguine}},\ }\bibfield  {title} {\enquote {\bibinfo {title}
  {{Anticipating a new physics signal in upcoming 21-cm power spectrum
  observations}},}\ }\href {\doibase 10.1103/PhysRevD.108.063503} {\bibfield
  {journal} {\bibinfo  {journal} {Phys. Rev. D}\ }\textbf {\bibinfo {volume}
  {108}},\ \bibinfo {pages} {063503} (\bibinfo {year} {2023})},\ \Eprint
  {http://arxiv.org/abs/2212.08082} {arXiv:2212.08082 [hep-ph]} \BibitemShut
  {NoStop}%
\bibitem [{\citenamefont {Barkana}\ \emph {et~al.}(2018)\citenamefont
  {Barkana}, \citenamefont {Outmezguine}, \citenamefont {Redigolo},\ and\
  \citenamefont {Volansky}}]{Barkana:2018qrx}%
  \BibitemOpen
  \bibfield  {author} {\bibinfo {author} {\bibfnamefont {Rennan}\ \bibnamefont
  {Barkana}}, \bibinfo {author} {\bibfnamefont {Nadav~Joseph}\ \bibnamefont
  {Outmezguine}}, \bibinfo {author} {\bibfnamefont {Diego}\ \bibnamefont
  {Redigolo}}, \ and\ \bibinfo {author} {\bibfnamefont {Tomer}\ \bibnamefont
  {Volansky}},\ }\bibfield  {title} {\enquote {\bibinfo {title} {{Strong
  constraints on light dark matter interpretation of the EDGES signal}},}\
  }\href {\doibase 10.1103/PhysRevD.98.103005} {\bibfield  {journal} {\bibinfo
  {journal} {Phys. Rev. D}\ }\textbf {\bibinfo {volume} {98}},\ \bibinfo
  {pages} {103005} (\bibinfo {year} {2018})},\ \Eprint
  {http://arxiv.org/abs/1803.03091} {arXiv:1803.03091 [hep-ph]} \BibitemShut
  {NoStop}%
\bibitem [{\citenamefont {Liu}\ \emph {et~al.}(2019)\citenamefont {Liu},
  \citenamefont {Outmezguine}, \citenamefont {Redigolo},\ and\ \citenamefont
  {Volansky}}]{Liu:2019knx}%
  \BibitemOpen
  \bibfield  {author} {\bibinfo {author} {\bibfnamefont {Hongwan}\ \bibnamefont
  {Liu}}, \bibinfo {author} {\bibfnamefont {Nadav~Joseph}\ \bibnamefont
  {Outmezguine}}, \bibinfo {author} {\bibfnamefont {Diego}\ \bibnamefont
  {Redigolo}}, \ and\ \bibinfo {author} {\bibfnamefont {Tomer}\ \bibnamefont
  {Volansky}},\ }\bibfield  {title} {\enquote {\bibinfo {title} {{Reviving
  Millicharged Dark Matter for 21-cm Cosmology}},}\ }\href {\doibase
  10.1103/PhysRevD.100.123011} {\bibfield  {journal} {\bibinfo  {journal}
  {Phys. Rev. D}\ }\textbf {\bibinfo {volume} {100}},\ \bibinfo {pages}
  {123011} (\bibinfo {year} {2019})},\ \Eprint
  {http://arxiv.org/abs/1908.06986} {arXiv:1908.06986 [hep-ph]} \BibitemShut
  {NoStop}%
\bibitem [{\citenamefont {Boddy}\ \emph {et~al.}(2018)\citenamefont {Boddy},
  \citenamefont {Gluscevic}, \citenamefont {Poulin}, \citenamefont {Kovetz},
  \citenamefont {Kamionkowski},\ and\ \citenamefont {Barkana}}]{Boddy:2018wzy}%
  \BibitemOpen
  \bibfield  {author} {\bibinfo {author} {\bibfnamefont {Kimberly~K.}\
  \bibnamefont {Boddy}}, \bibinfo {author} {\bibfnamefont {Vera}\ \bibnamefont
  {Gluscevic}}, \bibinfo {author} {\bibfnamefont {Vivian}\ \bibnamefont
  {Poulin}}, \bibinfo {author} {\bibfnamefont {Ely~D.}\ \bibnamefont {Kovetz}},
  \bibinfo {author} {\bibfnamefont {Marc}\ \bibnamefont {Kamionkowski}}, \ and\
  \bibinfo {author} {\bibfnamefont {Rennan}\ \bibnamefont {Barkana}},\
  }\bibfield  {title} {\enquote {\bibinfo {title} {{Critical assessment of CMB
  limits on dark matter-baryon scattering: New treatment of the relative bulk
  velocity}},}\ }\href {\doibase 10.1103/PhysRevD.98.123506} {\bibfield
  {journal} {\bibinfo  {journal} {Phys. Rev. D}\ }\textbf {\bibinfo {volume}
  {98}},\ \bibinfo {pages} {123506} (\bibinfo {year} {2018})},\ \Eprint
  {http://arxiv.org/abs/1808.00001} {arXiv:1808.00001 [astro-ph.CO]}
  \BibitemShut {NoStop}%
\bibitem [{\citenamefont {Alvey}\ \emph {et~al.}(2021)\citenamefont {Alvey},
  \citenamefont {Sabti}, \citenamefont {Tiki}, \citenamefont {Blas},
  \citenamefont {Bondarenko}, \citenamefont {Boyarsky}, \citenamefont
  {Escudero}, \citenamefont {Fairbairn}, \citenamefont {Orkney},\ and\
  \citenamefont {Read}}]{Alvey:2020xsk}%
  \BibitemOpen
  \bibfield  {author} {\bibinfo {author} {\bibfnamefont {James}\ \bibnamefont
  {Alvey}}, \bibinfo {author} {\bibfnamefont {Nashwan}\ \bibnamefont {Sabti}},
  \bibinfo {author} {\bibfnamefont {Victoria}\ \bibnamefont {Tiki}}, \bibinfo
  {author} {\bibfnamefont {Diego}\ \bibnamefont {Blas}}, \bibinfo {author}
  {\bibfnamefont {Kyrylo}\ \bibnamefont {Bondarenko}}, \bibinfo {author}
  {\bibfnamefont {Alexey}\ \bibnamefont {Boyarsky}}, \bibinfo {author}
  {\bibfnamefont {Miguel}\ \bibnamefont {Escudero}}, \bibinfo {author}
  {\bibfnamefont {Malcolm}\ \bibnamefont {Fairbairn}}, \bibinfo {author}
  {\bibfnamefont {Matthew}\ \bibnamefont {Orkney}}, \ and\ \bibinfo {author}
  {\bibfnamefont {Justin~I.}\ \bibnamefont {Read}},\ }\bibfield  {title}
  {\enquote {\bibinfo {title} {{New constraints on the mass of fermionic dark
  matter from dwarf spheroidal galaxies}},}\ }\href {\doibase
  10.1093/mnras/staa3640} {\bibfield  {journal} {\bibinfo  {journal} {Mon. Not.
  Roy. Astron. Soc.}\ }\textbf {\bibinfo {volume} {501}},\ \bibinfo {pages}
  {1188--1201} (\bibinfo {year} {2021})},\ \Eprint
  {http://arxiv.org/abs/2010.03572} {arXiv:2010.03572 [hep-ph]} \BibitemShut
  {NoStop}%
\bibitem [{\citenamefont {Bogorad}\ and\ \citenamefont
  {Toro}(2022)}]{Bogorad:2021uew}%
  \BibitemOpen
  \bibfield  {author} {\bibinfo {author} {\bibfnamefont {Zachary}\ \bibnamefont
  {Bogorad}}\ and\ \bibinfo {author} {\bibfnamefont {Natalia}\ \bibnamefont
  {Toro}},\ }\bibfield  {title} {\enquote {\bibinfo {title} {{Ultralight
  millicharged dark matter via misalignment}},}\ }\href {\doibase
  10.1007/JHEP07(2022)035} {\bibfield  {journal} {\bibinfo  {journal} {JHEP}\
  }\textbf {\bibinfo {volume} {07}},\ \bibinfo {pages} {035} (\bibinfo {year}
  {2022})},\ \Eprint {http://arxiv.org/abs/2112.11476} {arXiv:2112.11476
  [hep-ph]} \BibitemShut {NoStop}%
\bibitem [{\citenamefont {Fernandez}\ \emph {et~al.}(2022)\citenamefont
  {Fernandez}, \citenamefont {Kahn},\ and\ \citenamefont
  {Shelton}}]{Fernandez:2021iti}%
  \BibitemOpen
  \bibfield  {author} {\bibinfo {author} {\bibfnamefont {Nicolas}\ \bibnamefont
  {Fernandez}}, \bibinfo {author} {\bibfnamefont {Yonatan}\ \bibnamefont
  {Kahn}}, \ and\ \bibinfo {author} {\bibfnamefont {Jessie}\ \bibnamefont
  {Shelton}},\ }\bibfield  {title} {\enquote {\bibinfo {title} {{Freeze-in,
  glaciation, and UV sensitivity from light mediators}},}\ }\href {\doibase
  10.1007/JHEP07(2022)044} {\bibfield  {journal} {\bibinfo  {journal} {JHEP}\
  }\textbf {\bibinfo {volume} {07}},\ \bibinfo {pages} {044} (\bibinfo {year}
  {2022})},\ \Eprint {http://arxiv.org/abs/2111.13709} {arXiv:2111.13709
  [hep-ph]} \BibitemShut {NoStop}%
\bibitem [{\citenamefont {Harris}\ \emph {et~al.}(2020)\citenamefont {Harris},
  \citenamefont {Millman}, \citenamefont {Van Der~Walt}, \citenamefont
  {Gommers}, \citenamefont {Virtanen}, \citenamefont {Cournapeau},
  \citenamefont {Wieser}, \citenamefont {Taylor}, \citenamefont {Berg},
  \citenamefont {Smith} \emph {et~al.}}]{harris2020array}%
  \BibitemOpen
  \bibfield  {author} {\bibinfo {author} {\bibfnamefont {Charles~R}\
  \bibnamefont {Harris}}, \bibinfo {author} {\bibfnamefont {K~Jarrod}\
  \bibnamefont {Millman}}, \bibinfo {author} {\bibfnamefont {St{\'e}fan~J}\
  \bibnamefont {Van Der~Walt}}, \bibinfo {author} {\bibfnamefont {Ralf}\
  \bibnamefont {Gommers}}, \bibinfo {author} {\bibfnamefont {Pauli}\
  \bibnamefont {Virtanen}}, \bibinfo {author} {\bibfnamefont {David}\
  \bibnamefont {Cournapeau}}, \bibinfo {author} {\bibfnamefont {Eric}\
  \bibnamefont {Wieser}}, \bibinfo {author} {\bibfnamefont {Julian}\
  \bibnamefont {Taylor}}, \bibinfo {author} {\bibfnamefont {Sebastian}\
  \bibnamefont {Berg}}, \bibinfo {author} {\bibfnamefont {Nathaniel~J}\
  \bibnamefont {Smith}},  \emph {et~al.},\ }\bibfield  {title} {\enquote
  {\bibinfo {title} {Array programming with numpy},}\ }\href@noop {} {\bibfield
   {journal} {\bibinfo  {journal} {Nature}\ }\textbf {\bibinfo {volume}
  {585}},\ \bibinfo {pages} {357--362} (\bibinfo {year} {2020})}\BibitemShut
  {NoStop}%
\bibitem [{\citenamefont {Virtanen}\ \emph {et~al.}(2020)\citenamefont
  {Virtanen}, \citenamefont {Gommers}, \citenamefont {Oliphant}, \citenamefont
  {Haberland}, \citenamefont {Reddy}, \citenamefont {Cournapeau}, \citenamefont
  {Burovski}, \citenamefont {Peterson}, \citenamefont {Weckesser},
  \citenamefont {Bright} \emph {et~al.}}]{virtanen2020scipy}%
  \BibitemOpen
  \bibfield  {author} {\bibinfo {author} {\bibfnamefont {Pauli}\ \bibnamefont
  {Virtanen}}, \bibinfo {author} {\bibfnamefont {Ralf}\ \bibnamefont
  {Gommers}}, \bibinfo {author} {\bibfnamefont {Travis~E}\ \bibnamefont
  {Oliphant}}, \bibinfo {author} {\bibfnamefont {Matt}\ \bibnamefont
  {Haberland}}, \bibinfo {author} {\bibfnamefont {Tyler}\ \bibnamefont
  {Reddy}}, \bibinfo {author} {\bibfnamefont {David}\ \bibnamefont
  {Cournapeau}}, \bibinfo {author} {\bibfnamefont {Evgeni}\ \bibnamefont
  {Burovski}}, \bibinfo {author} {\bibfnamefont {Pearu}\ \bibnamefont
  {Peterson}}, \bibinfo {author} {\bibfnamefont {Warren}\ \bibnamefont
  {Weckesser}}, \bibinfo {author} {\bibfnamefont {Jonathan}\ \bibnamefont
  {Bright}},  \emph {et~al.},\ }\bibfield  {title} {\enquote {\bibinfo {title}
  {Scipy 1.0: fundamental algorithms for scientific computing in python},}\
  }\href@noop {} {\bibfield  {journal} {\bibinfo  {journal} {Nature methods}\
  }\textbf {\bibinfo {volume} {17}},\ \bibinfo {pages} {261--272} (\bibinfo
  {year} {2020})}\BibitemShut {NoStop}%
\bibitem [{\citenamefont {Hunter}(2007)}]{hunter2007matplotlib}%
  \BibitemOpen
  \bibfield  {author} {\bibinfo {author} {\bibfnamefont {John~D}\ \bibnamefont
  {Hunter}},\ }\bibfield  {title} {\enquote {\bibinfo {title} {Matplotlib: A 2d
  graphics environment},}\ }\href@noop {} {\bibfield  {journal} {\bibinfo
  {journal} {Computing in science \& engineering}\ }\textbf {\bibinfo {volume}
  {9}},\ \bibinfo {pages} {90--95} (\bibinfo {year} {2007})}\BibitemShut
  {NoStop}%
\bibitem [{\citenamefont {Rohatgi}(2018)}]{rohatgi2018webplotdigitizer}%
  \BibitemOpen
  \bibfield  {author} {\bibinfo {author} {\bibfnamefont {A}~\bibnamefont
  {Rohatgi}},\ }\bibfield  {title} {\enquote {\bibinfo {title}
  {Webplotdigitizer},}\ }\href@noop {} {\  (\bibinfo {year}
  {2018})}\BibitemShut {NoStop}%
\bibitem [{\citenamefont {et~al.}(2016)}]{soton403913}%
  \BibitemOpen
  \bibfield  {author} {\bibinfo {author} {\bibfnamefont {Thomas~Kluyver}\
  \bibnamefont {et~al.}},\ }\bibfield  {title} {\enquote {\bibinfo {title}
  {Jupyter notebooks a publishing format for reproducible computational
  workflows},}\ }in\ \href {https://eprints.soton.ac.uk/403913/} {\emph
  {\bibinfo {booktitle} {Positioning and Power in Academic Publishing: Players,
  Agents and Agendas}}}\ (\bibinfo  {publisher} {IOS Press},\ \bibinfo {year}
  {2016})\ pp.\ \bibinfo {pages} {87--90}\BibitemShut {NoStop}%
\bibitem [{\citenamefont {Wolfram}(2003)}]{wolfram2003mathematica}%
  \BibitemOpen
  \bibfield  {author} {\bibinfo {author} {\bibfnamefont {Stephen}\ \bibnamefont
  {Wolfram}},\ }\href@noop {} {\emph {\bibinfo {title} {The mathematica
  book}}}\ (\bibinfo  {publisher} {Wolfram Research, Inc.},\ \bibinfo {year}
  {2003})\BibitemShut {NoStop}%
\end{thebibliography}%
\end{document}